\pdfoutput=1
\documentclass[sigconf,nonacm]{acmart}

\usepackage{booktabs} 

\copyrightyear{2019}
\acmYear{2019} 
\setcopyright{iw3c2w3}
\acmConference[WWW '19 Companion]{Companion Proceedings of the 2019 World Wide Web Conference}{May 13--17, 2019}{San Francisco, CA, USA}
\acmBooktitle{Companion Proceedings of the 2019 World Wide Web Conference (WWW '19 Companion), May 13--17, 2019, San Francisco, CA, USA}
\acmPrice{}
\acmDOI{10.1145/3308560.3316699}
\acmISBN{978-1-4503-6675-5/19/05}

\usepackage{balance}

\usepackage{natbib}
\usepackage{amsmath}
\usepackage{graphicx}
\usepackage{epstopdf}
\graphicspath{{./graphics/}}
\usepackage{subfig}
\usepackage{multirow}
\usepackage{siunitx}
\usepackage{makecell}
\usepackage[margin=false,inline=true, draft]{fixme}
\usepackage{enumitem}  
\usepackage{hyperref}

\DeclareMathOperator{\score}{score}
\DeclareMathOperator{\PDCN}{PD-CN}
\DeclareMathOperator{\PDAA}{PD-AA}

\begin{document}

\title[Personalized Degrees: Effects on Link Formation in Dynamic Networks from an Egocentric Perspective]
{Personalized Degrees: Effects on Link Formation in Dynamic Networks from an Egocentric Perspective}

\author{Makan Arastuie}
\affiliation{%
 \institution{University of Toledo}
 \city{Toledo}
 \state{OH}
 \country{USA}
}
\email{makan.arastuie@rockets.utoledo.edu}

\author{Kevin S. Xu}
\affiliation{%
 \institution{University of Toledo}
 \city{Toledo}
 \state{OH}
 \country{USA}
}
\email{kevin.xu@utoledo.edu}

\begin{abstract}
Understanding mechanisms driving link formation in dynamic social networks is a long-standing problem that has implications to understanding social structure as well as link prediction and recommendation. 
Social networks exhibit a high degree of transitivity, which explains the successes of common neighbor-based methods for link prediction. 
In this paper, we examine mechanisms behind link formation from the perspective of an ego node.
We introduce the notion of \emph{personalized degree} for each neighbor node of the ego, which is the number of other neighbors a particular neighbor is connected to. 
From empirical analyses on four on-line social network datasets, we find that 
neighbors with higher personalized degree are \emph{more likely} to lead to new link formations when they serve as common neighbors with other nodes, both in undirected and directed settings. 
This is complementary to the finding of Adamic and Adar \citep{adamic2003friends} that neighbor nodes with higher (global) degree are \emph{less likely} to lead to new link formations. 
Furthermore, on directed networks, we find that personalized out-degree has a stronger effect on link formation than personalized in-degree, whereas global in-degree has a stronger effect than global out-degree. 
We validate our empirical findings through several link recommendation experiments and observe that incorporating both personalized and global degree into link recommendation greatly improves accuracy.
\end{abstract}

%
%



\maketitle

\section{Introduction}
\label{sec:intro}
Ever since the rise of different social media platforms, the need for analyzing complex social networks has been growing at an exponential rate. A fundamental problem when analyzing social network data is to understand mechanisms driving link formation. Specifically, in the absence of additional information about the nodes in the network, what properties of the network itself tend to be associated with the formation of future links? Such properties could then be incorporated into models for the evolution of social networks over time \citep{Leskovec2008}, which can then be used to predict the formation of links in the future. Predicting future or missing links in a network has been a major area of research and was formalized by \citet{liben2007link} as the link prediction problem. It has been widely studied since then, and many different approaches and algorithms have been proposed; we refer readers to \citet{Lu2011} for a survey of the literature. 

General principles that govern the behavior of network dynamics form the core of many link prediction and recommendation problems. Common neighbors \citep{newman2001clustering} and Adamic/Adar \citep{adamic2003friends} are two widely used principles which both predate much of the research on link prediction. 

\citet{adamic2003friends} found that the ``popularity'' or global degree\footnote{In this paper we will refer to the degree of a node, i.e.~the number of nodes a particular node it is connected to, as \emph{global degree} to avoid confusion with personalized degree.} of a common neighbor of a pair of nodes has an inverse relationship with the likelihood of the two nodes forming an edge in the future. 
In this paper, we introduce a new principle that is observed to be present in the dynamics of social networks, the \emph{personalized degree} of common neighbors, and examine its behavior and how it correlates with the formation of future links from an egocentric perspective.

This paper is divided up into three parts. In Section \ref{sec:personalized-degree}, we introduce personalized degree and examine its distribution over several datasets.
Section \ref{sec:empirical} consists of a series of empirical analyses on the effects of personalized degree on future link formation. 
In Section \ref{sec:link-recommendation}, we validate our empirical findings by incorporating them into two link recommendation algorithms and examining the effects of edge directionality on link recommendation.

Our main contributions are as follows:
\begin{itemize}
\item We find that the personalized degree of a neighbor behaves in the opposite manner compared to its global degree; that is, common  neighbors with higher personalized degree are \emph{more likely} to be predictive of  future links, both in directed and undirected networks.

\item In directed networks, we find that personalized out-degree has a stronger effect than personalized in-degree and that global in-degree has a stronger effect than global out-degree. Consequently, limiting personalized and global degree to personalized out-degree and global in-degree, respectively, provides more accurate predictions of link formations.

\item We validate the previous two findings through link recommendation experiments, where we find that incorporating personalized degree alone improves mean link prediction accuracy by $2\%$ to $30\%$ and incorporating both personalized and global degrees improves accuracy by $6\%$ to $35\%$. Incorporating directions of edges further improves accuracy by up to $11\%$.

\end{itemize}

\section{Background} \label{sec:background}

\subsection{Link Prediction}
Information in networks is carried out in links between nodes. The presence or absence of a link has implications on the social structure. The importance of links makes the concept of predicting new or missing links in a network extremely desirable. This concept is formalized as the link prediction problem \citep{liben2007link} and is widely used in a variety of applications. A few examples include recommendation of people to follow in on-line social media networks \citep{Gupta2013}, imputation of missing links from partially observed networks \citep{Kossinets2006a,Lu2011}, and validation of models for network formation \citep{Leskovec2008}.

All link prediction methods, as mentioned in \cite{liben2007link}, assign an estimate $\score(x, y)$ to all pairs of nodes $(x, y)$ without any links. All scores are then ranked in decreasing order. The higher the rank, the higher the probability of existing an edge between that pair of node. The earliest and simplest methods for link prediction are based on node neighborhoods, where two nodes $x$ and $y$ are predicted to be more likely to form a link in the future if their sets of neighbors $\Gamma(x)$ and $\Gamma(y)$ have large overlap \citep{liben2007link}. 

The simplest method, typically referred to as just common neighbors (CN), simply uses the number of common neighbors as the predicted score:
\begin{equation} 
	\label{eq:cn-score}
	\score(x, y) := |\Gamma(x) \cap \Gamma(y)|.
\end{equation}
Another method, commonly referred to as Adamic/Adar (AA), sums over all common neighbors, but weights the common neighbors with a lower global degree more heavily in order to indicate that they are more predictive than the ones with a higher global degree \citep{adamic2003friends}: 
\begin{equation}
	\label{eq:aa-score}
	\score(x, y) := \sum_{z\in \Gamma(x) \cap \Gamma(y)} {\left(\log{|\Gamma(z)|}\right)^{-1}}
\end{equation}
The Adamic/Adar link predictor has generally been found to be more accurate than common neighbors on a variety of network datasets \citep{liben2007link}. 

\subsection{Egocentric Perspective \& Link Recommendation}

Link prediction is usually considered at the \emph{global} network level, where the objective is to predict the most likely links to be formed between any pair of nodes in the network that do not already have a link.
Moreover, it has also been done at the level of an ego node, where the objective is to predict the most likely links to be formed involving the ego node. This problem is commonly referred to as \emph{link recommendation} \citep{Yin2010}, because the predicted nodes are often used as recommendations for the ego node, e.g.~the Who to Follow feature in Twitter \citep{Gupta2013}.

In this study we take an egocentric perspective. Throughout most of our analyses, we first select a node to be the ego and its neighbors to be the common neighbors. Next, we examine link formation between the ego and all neighbors of the common neighbors that are not common neighbors themselves; or simply put, nodes that are 2 hops away from the ego node. This is specially important since most links are formed between nodes that are 2 hops away due to the locality of link formation \citep{Yang2015}.

\subsection{Related Work}
\label{sec:related}

Many link prediction methods have also attempted to incorporate mechanisms behind link formation into link prediction. \citet{Leskovec2008} analyze the microscopic evolution of social networks and incorporate their findings on node and link duration as well as triangle closing into a network evolution model that can be used to predict future links. \citet{Liu2013c} investigate the role of community structure on link formation and propose a link prediction algorithm based on a model for community detection. \citet{cannistraci2013link} introduced the concept of local-community-paradigm and suggested that two nodes are more likely to link together if their common neighbors are highly connected within themselves. They referred to the edges between common neighbors as local-community-links (LCL). \citet{wu2016link}, inspired by LCL and the fact that links between common neighbors of two nodes form a triangle, claimed that the probability of two nodes forming an edge has a direct relationship with the number of triangles passing through their common neighbors and an inverse relationship with their global degree, similar to AA. 
To the best of our knowledge, these types of studies have usually taken place at the global network level, not from the egocentric perspective that we consider.

\subsection{Datasets}
\label{sec:datasets}

\begin{table}[t]
  \centering
  \caption{Summary statistics of datasets used in this paper.}
  \label{tab:dataset-info}
  \begin{tabular}{lcccc c}
    \hline
    \textbf{Dataset} & \textbf{Facebook} & \textbf{Google+} & \textbf{Flickr} & \textbf{Digg} \\
    \hline 
    \textbf{Directed?} & No & Yes & Yes & Yes \\
    \textbf{\# of Nodes} & 63,731 & $\sim$29M & $\sim$2.3M & 63,740 \\
    \textbf{\# of Edges} & 817,090 & $\sim$462M & $\sim$33M & 837,104 \\
    \textbf{\# of Snapshots} & 10 & 4 & 5 & 16 \\
    \textbf{Duration (days)} & $\sim$869 & $\sim$98 & $\sim$198 & $\sim$1431 \\
    \hline
  \end{tabular}
\end{table}

In this study, we analyze three directed and one undirected on-line social networks, as shown in Table \ref{tab:dataset-info}. In all four networks, nodes represent users of that platform. 

\subsubsection{Facebook}
The Facebook New Orleans friendship network is an undirected network, collected by \citet{viswanath2009evolution}, in which every link represents a friendship between two nodes with a time-stamp of when the friendship was formed.
The data trace covers the period between September 2006 and January 2009, and each snapshot lasts for 90 days.

Around March 2008, Facebook introduced the "People You May Know" (PYMK) feature to recommend new friends to their users. While link formations prior to the introduction of PYMK were likely the result of an organic process, the PYMK feature is likely to influence the link formation process. Thus, we divided the entire Facebook dataset into two sets, before and after PYMK. The first 6 extracted snapshots consist entirely of link formations before PYMK, and the remaining 4 came after PYMK.

\subsubsection{Google+}
\label{sec:data-google}
This dataset was crawled from the Google+ network, starting from July 6 until October 11, 2011, by \citet{gong2012evolution}, and it covers more than 70\% of the entire Google+ network at that time.
Incoming and outgoing friends are represented by directed links which belong to one of the 4 predefined snapshots.

Due to the large network size, we randomly selected about 500,000 nodes that appeared in the first snapshot of the network to serve as ego nodes in all of our analyses. The same set of randomly selected nodes are used in all experiments for consistency. Furthermore, if an ego node had more than 100,000 nodes that were at most 2 hops away, then it was not used in the experiments.

\subsubsection{Flickr}
\label{sec:data-flickr}
Flickr is a media hosting service where users can host images and videos and share them with their friends.
Most snapshots in the Flickr dataset \citep{mislove-2008-flickr} are 30 days long. 
Moreover, in order to manage the large size of this network, we made the exact same random sampling decisions as we did for Google+.

\subsubsection{Digg}
\label{sec:data-digg}
Digg is a news aggregator where users can follow each other, share their own news or blog posts, and up/down vote others' posts.
We divided the Digg friendship dataset \cite{hogg2012social} into 90-day long snapshots, also ignored all self-loops, inactive users, and users with an out-degree of zero.

\section{Personalized Degree} \label{sec:personalized-degree}

\begin{figure}[t]
\centering
\includegraphics[scale=0.4]{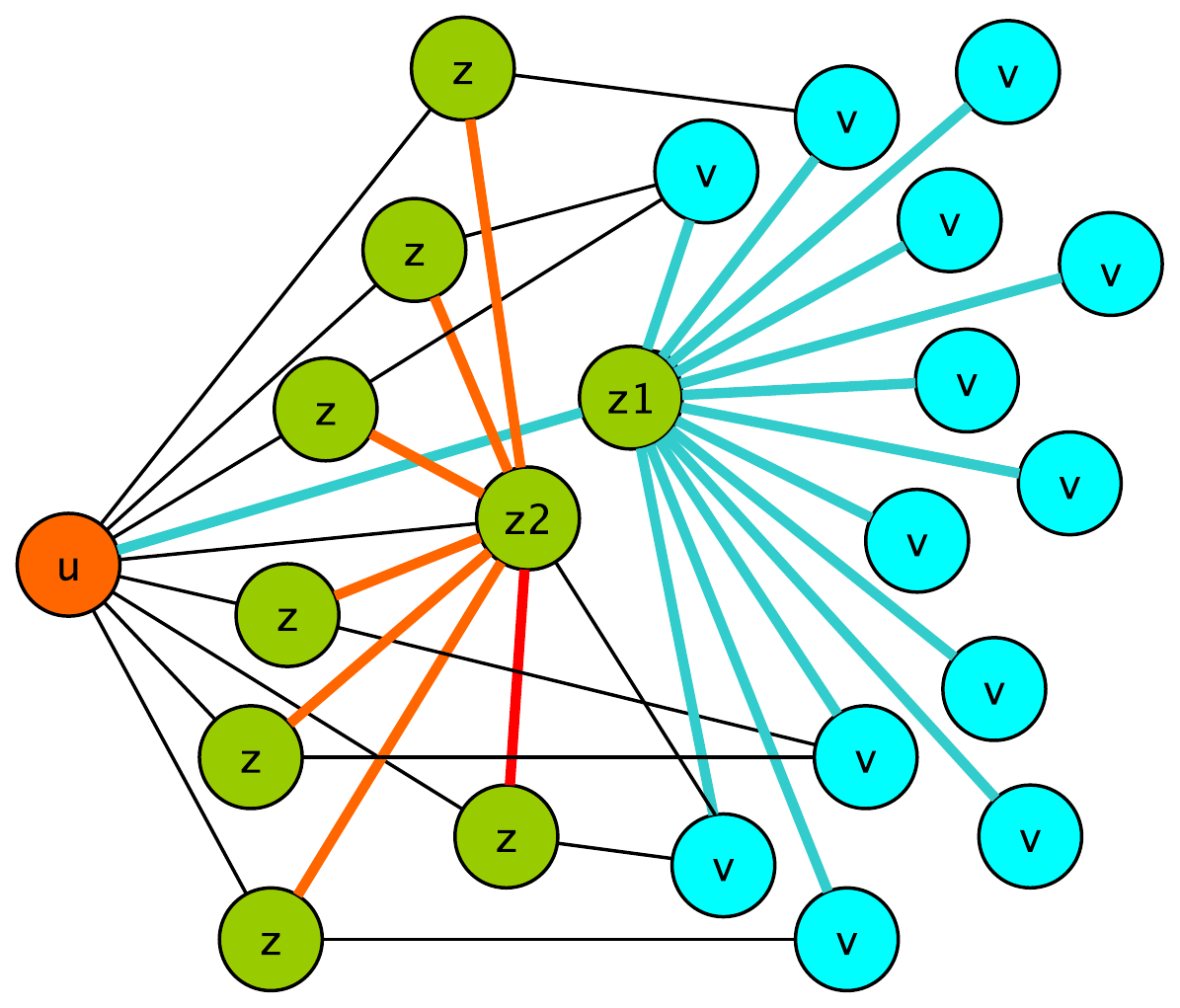} \\
Neighbor node z1: Global degree 14, personalized degree 0 \\
Neighbor node z2: Global degree 9, personalized degree 7
\caption{Example of a high global and a high personalized degree in a 2-hop egocentric network.}
\label{fig:ex-high-local-vs-global}
\end{figure}

For any given ego node, we define the \emph{personalized degree} of a neighbor node to be the number of other nodes that have links with both the ego and the neighbor node.
Thus, while each node in a network has a single global degree, it has a \emph{different} personalized degree with respect to every ego node. 
As shown by node z1 in Figure \ref{fig:ex-high-local-vs-global}, a neighbor node may have a high global degree, but a low personalized degree. Moreover, personalized degree of a neighbor node is always at least 1 less than its global degree and at least 2 less in order to be a common neighbor between the ego node and a node 2 hops away from the ego. 

In directed networks, just like global degree, personalized degree is divided into personalized in-degree and personalized out-degree. 
These respectively refer to the number of predecessors and successors of the neighbor node that also have an edge with the ego, where the direction of the edge between the ego and all other nodes, including the neighbor node, is the same.

It is important not to confuse personalized degree with localized clustering coefficient, first introduced by \citet{watts1998collective}. Given a node, localized clustering coefficient measures how close its neighbors are to forming a clique. In measuring local clustering coefficient, we fix one node and look at the number of its formed triangles over the total possible number of triangles with its neighbors. In personalized degree, every neighbor of a neighbor node does form a triangle, however, we not only consider the total number of possible triangles, but we also fix two nodes of each triangle, the ego and the common neighbor, instead of only one fixed node.

\subsection{Personalized Degree Distribution} \label{sec:personalized-degree-dist}
First, we examine both directed and undirected personalized degree distributions to better understand their behavior.
We consider each node in the network to be an ego and compute the personalized degree of all of its neighbors. 
We then plot the distribution of all computed personalized degrees on a log-log scale, similar to plotting the global degree distribution of a network. In addition, in directed networks, in order to stay consistent with our egocentric perspective, we consider all successors of an ego to be its neighbors. Keep in mind that these neighbors may or may not be the predecessors of the ego. Moreover, if it is possible for a node degree to be zero, to observe these instances on a log-log scale, we shift all node degrees up by one.
Due to the large number of nodes in all analyzed networks, we plot degree distributions using uniform log-binned scatter plots.

\begin{figure}[t]
  \centering
  \subfloat[Before PYMK]{
  \includegraphics[width=0.48\columnwidth]{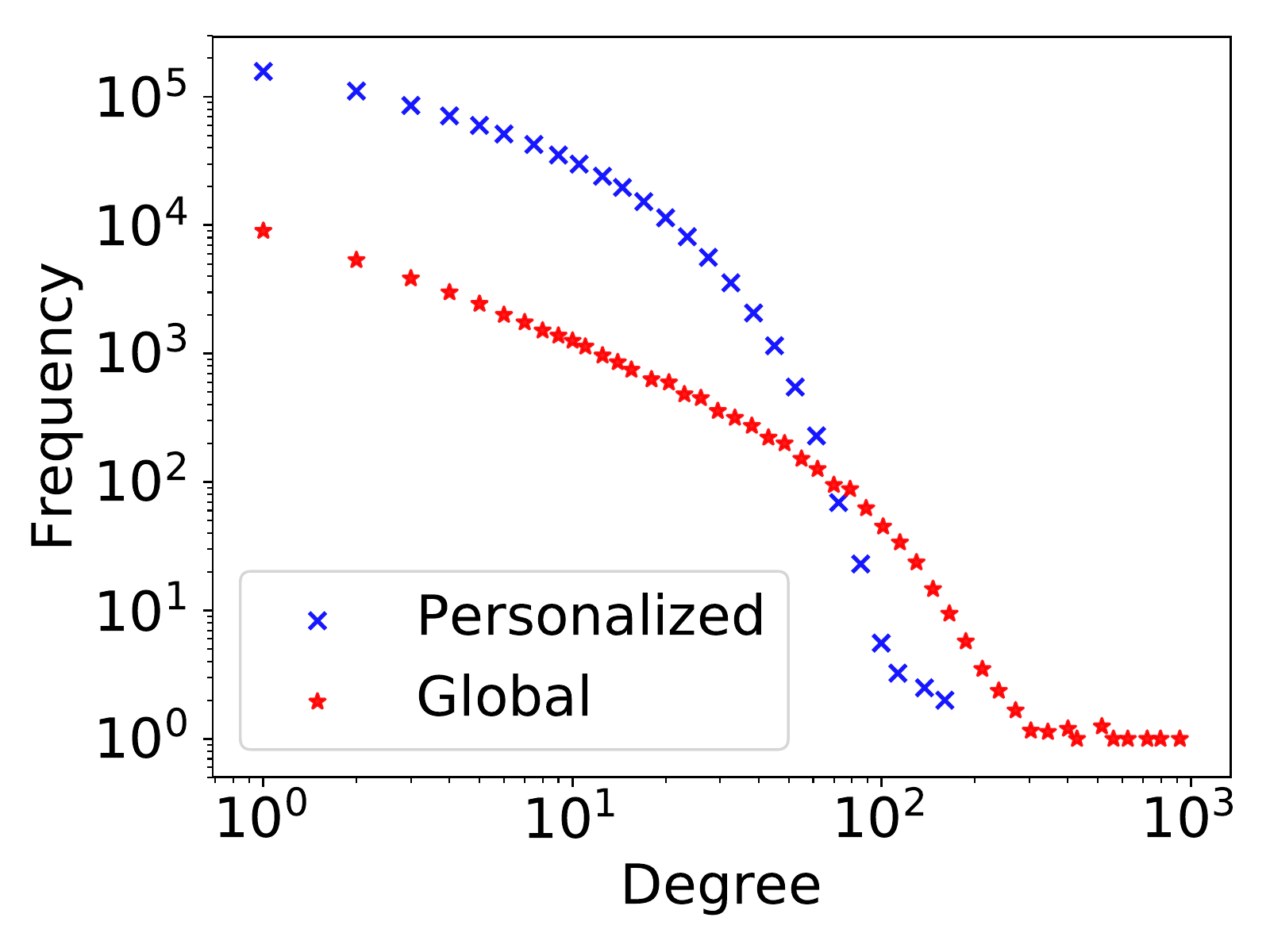}
  \label{fig:deg-dist-fb-before-pymk}}
  \subfloat[After PYMK]{
  \includegraphics[width=0.48\columnwidth]{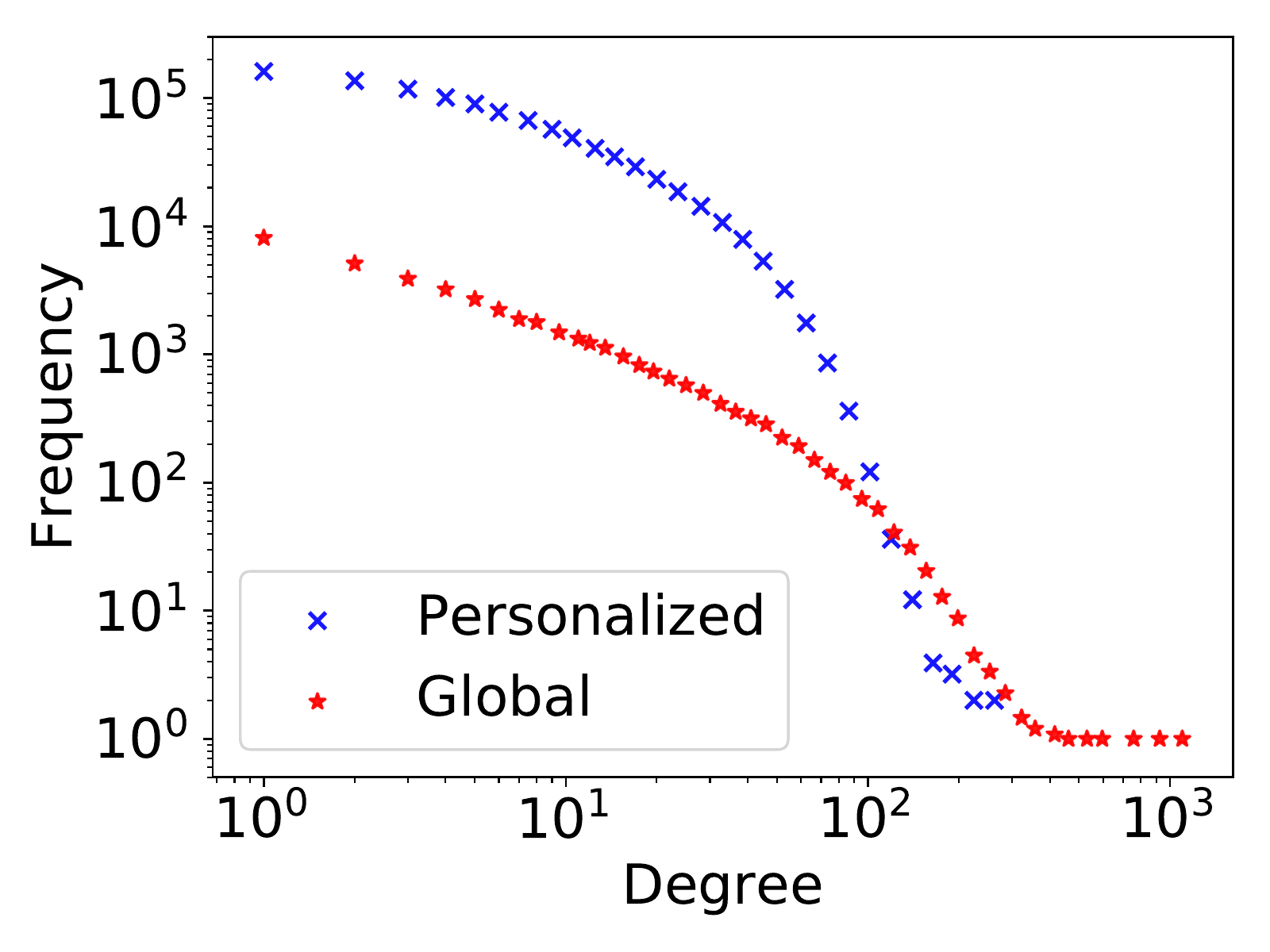}
  \label{fig:deg-dist-fb-after-pymk}}
  \caption{Facebook's global and personalized degree distributions before and after the introduction of the People You May Know (PYMK) feature.}
  \label{fig:deg-dist-fb}
\end{figure}

\begin{figure}[t]
  \centering
  \includegraphics[width=0.48\columnwidth]{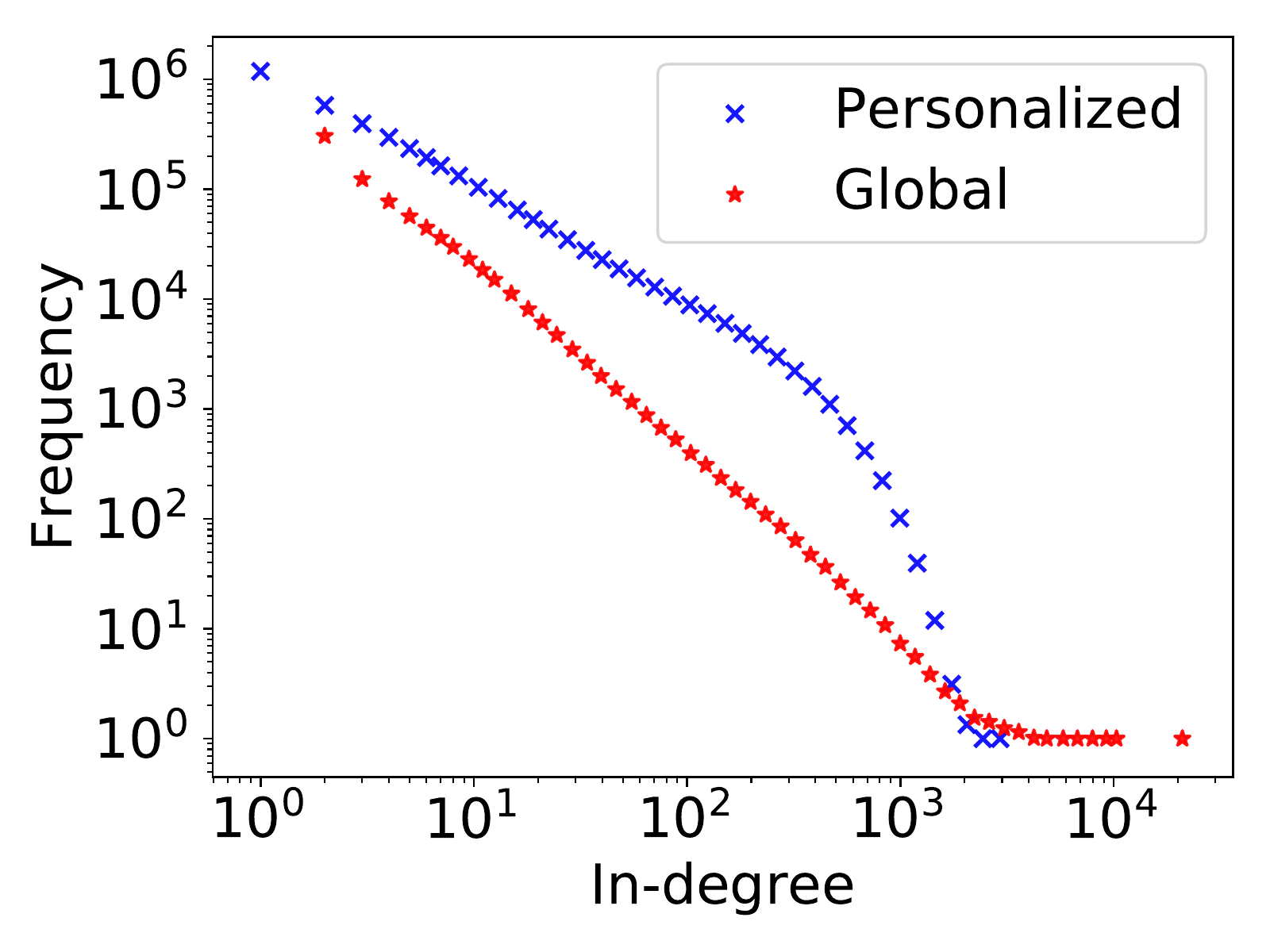}
  \includegraphics[width=0.48\columnwidth]{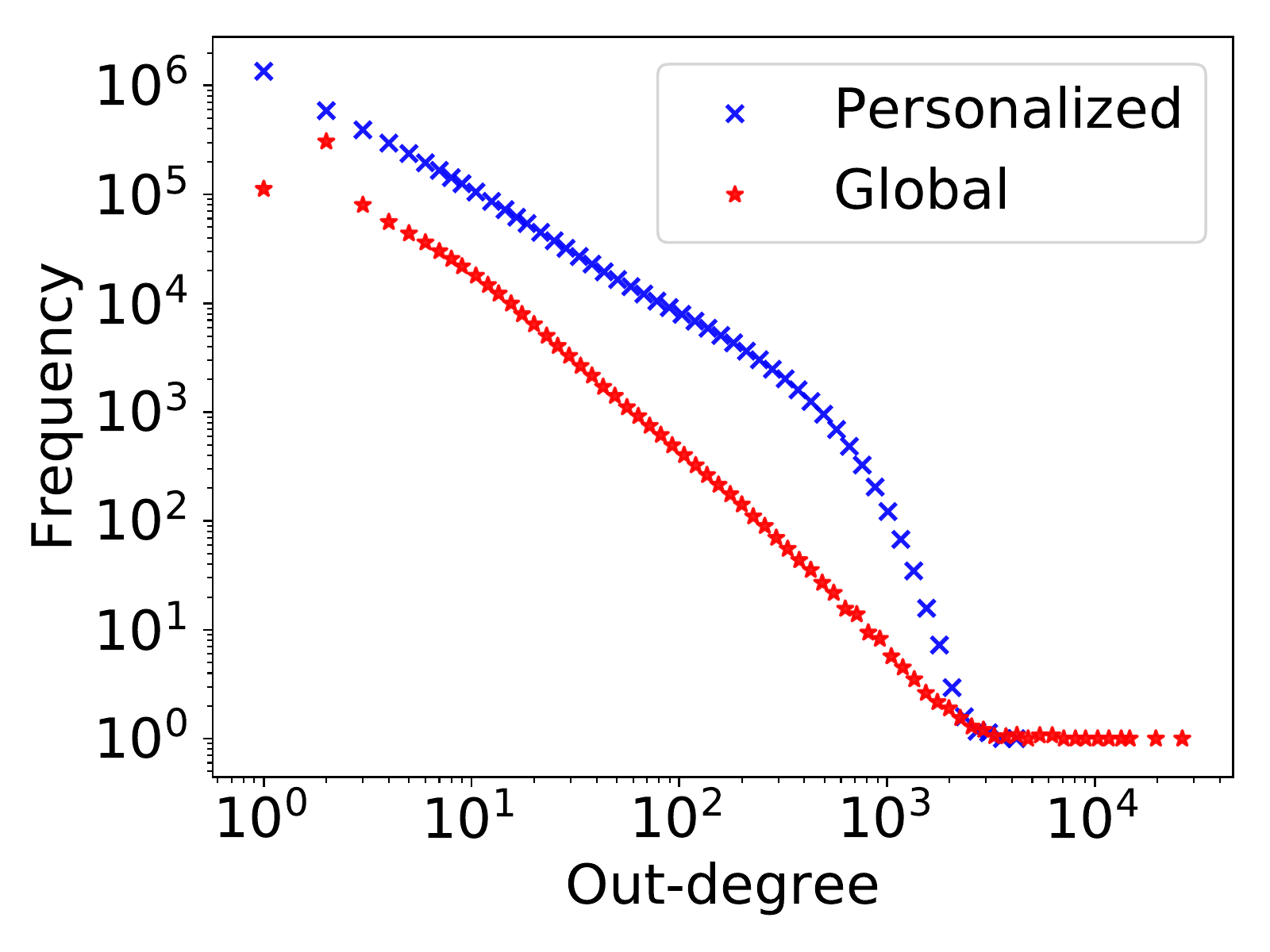}
  \caption{Flickr's global and personalized in-degree and out-degree distributions.}
  \label{fig:deg-dist-flickr}
\end{figure}

A widely accepted claim in modern network science is that most complex networks, including social networks, are scale-free \citep{barabasi1999emergence, barabasi2003scale, mitzenmacher2004brief}, meaning that their degree distribution follow a power law. As it is shown in Figures \ref{fig:deg-dist-fb} and \ref{fig:deg-dist-flickr}, the global degree distribution of both Facebook and Flickr seem to follow a power law as well; nevertheless, Flickr seem to be more strongly scale-free compared to Facebook.
Furthermore, both personalized degree distributions seem to follow a power law distribution; however, they both have lighter tails than a normal power law. Due to the light-tailed characteristic of these distribution, it is more fitting to describe their distribution by a truncated power law rather than a regular heavy-tailed power law distribution. Degree distributions of both Google+ and Digg are very similar to that of Flickr.

\section{Empirical Observations} \label{sec:empirical}
In this section, we perform an empirical analysis on the role of common neighbors' degrees, both global and personalized, on link formation from an egocentric perspective on all four datasets.

\subsection{Empirical Analysis Procedure} \label{sec:empirical-procedure}

To investigate the role of neighbor node degrees on link formation from an egocentric perspective, we consider every node in the network to be an ego node, and then we analyze the possibility of link formation between the ego node and all nodes which are 2 hops away from it, across all snapshots.
More specifically, for every network, beginning with the first snapshot, we consider every node to be an ego node, then for each node $v$ that is 2 hops away from the ego, we check whether it formed an edge with the ego node in the next snapshot. We separate the set of nodes that formed edges with the ego node, denoted by $V_f$, from the set of nodes that didn't, denoted by $V_{nf}$.

For each node $v_f \in V_f$, we calculate the mean log global and mean log personalized degrees of all common neighbors between $v_f$ and the ego node. As we observed in Section \ref{sec:personalized-degree-dist}, both degrees follow either a truncated or a regular power law distribution, thus simply taking the mean of global and personalized degrees will not result in a reliable summary statistic. Thereby, we took the log of both global and personalized degrees to adjust for this behavior.

Next we take the average of mean log global and mean log personalized degree over all nodes in $V_f$. We repeat the same process for the nodes $v_{nf} \in V_{nf}$ that did not form an edge with the ego node. Finally, we take the mean over all snapshots of the network.

The aforementioned process is repeated for, and averaged over, all ego nodes. We then compare the mean log global and mean log personalized degrees for all nodes which formed an edge with an ego to the ones that did not. 

Note that in order to make sure the averages for both formed and not-formed groups are over the exact same ego nodes in all snapshots, for every ego node, we excluded snapshots where either formed or not-formed group was empty. Therefore, if all snapshots of an ego node were excluded, the ego node was naturally removed from the analysis.

\subsection{Observations on Facebook Data} \label{sec:empirical-facebook}
\begin{figure}[t]
  \centering
  \subfloat[Global degree]{
  \includegraphics[width=0.48\columnwidth]{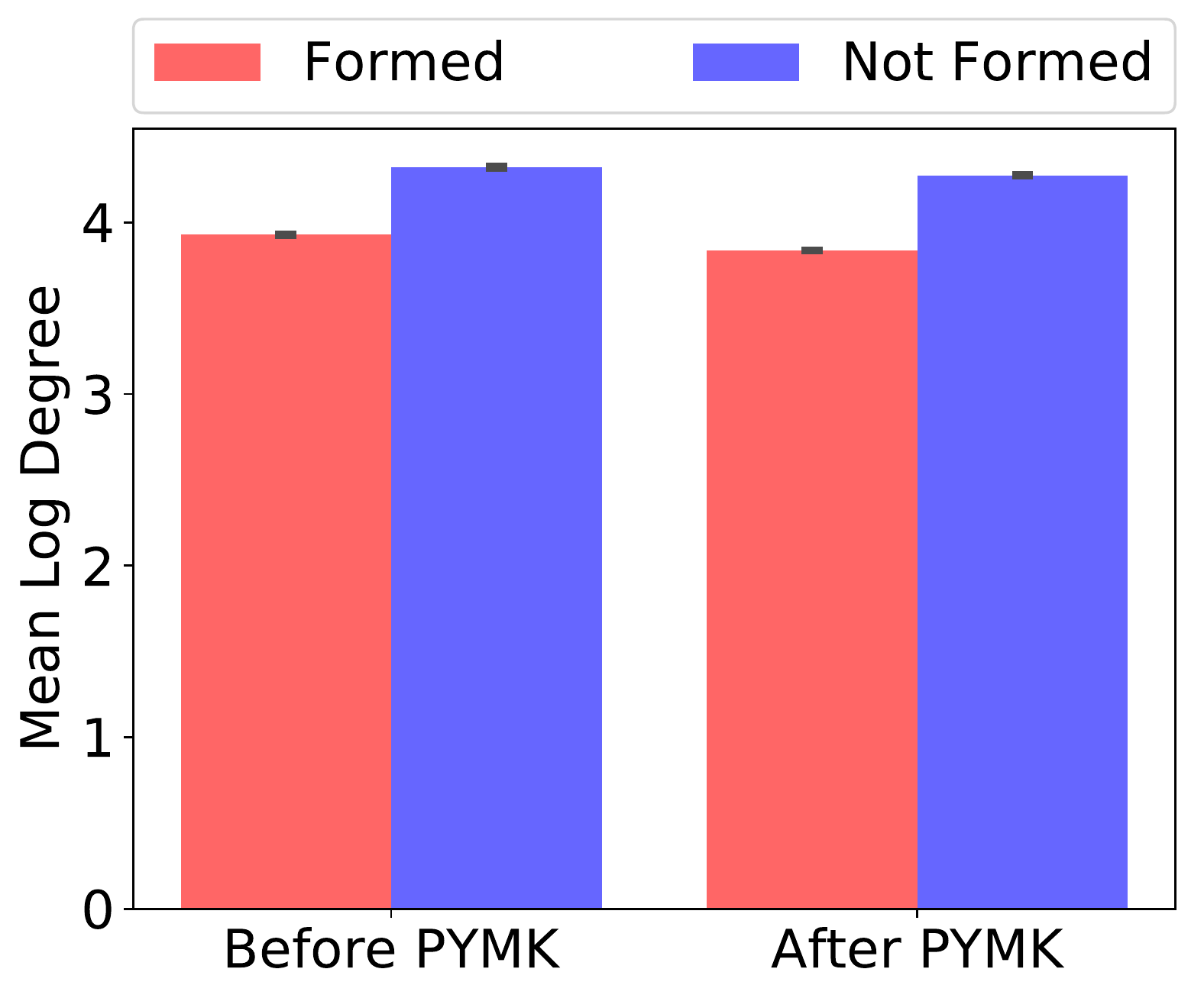}
  \label{fig:empirical-fb-mean-global}}
  \subfloat[Personalized degree]{
  \includegraphics[width=0.48\columnwidth]{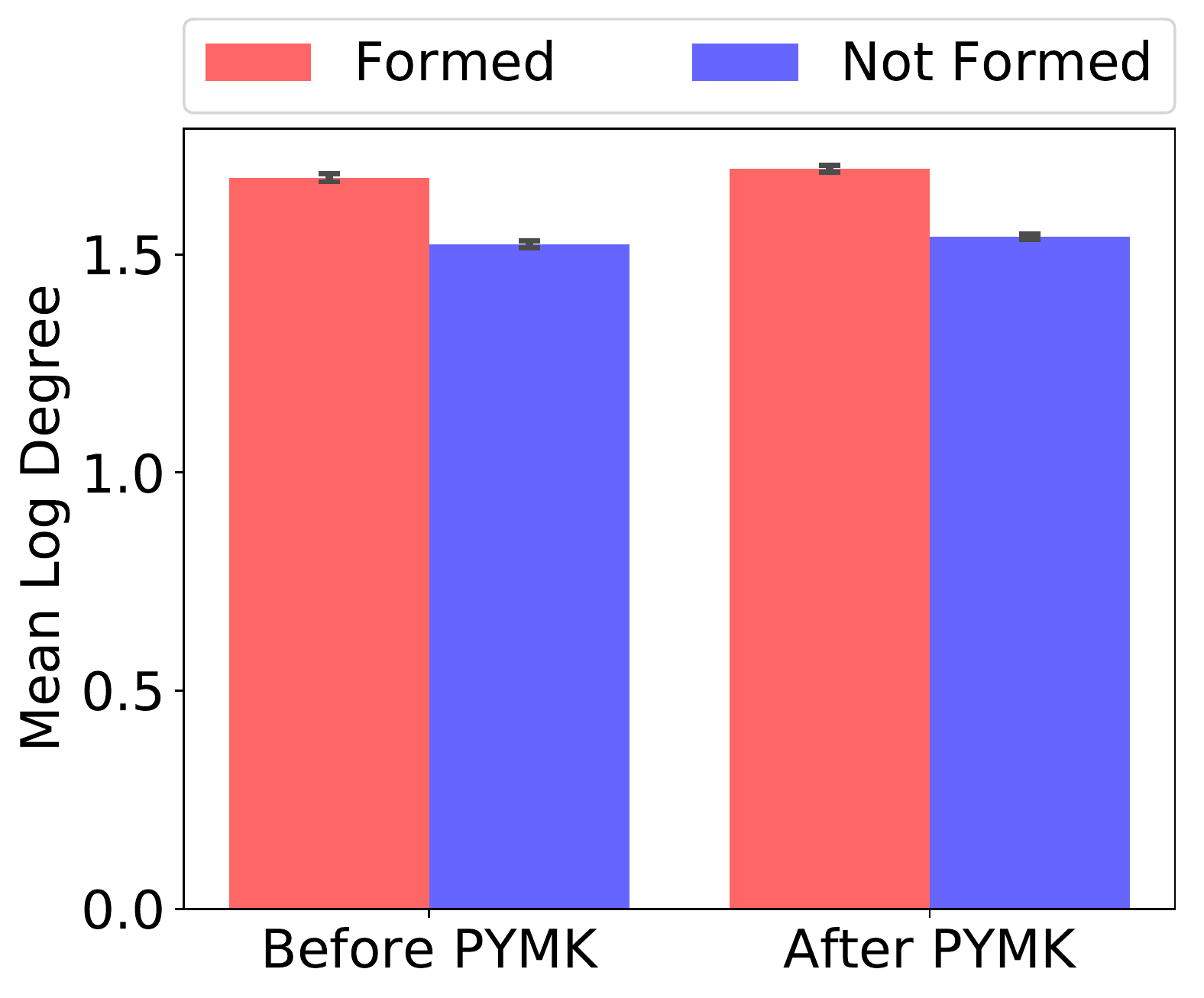}
  \label{fig:empirical-fb-mean-local}}
  \caption{Mean log global and mean log personalized degrees ($\pm$2 standard errors) of common neighbors in Facebook data before PYMK (19,162 ego nodes) and after PYMK (39,282 ego nodes).
  }
  \label{fig:empirical-fb-mean}
\end{figure}

As shown in Figure \ref{fig:empirical-fb-mean-global}, we observe that, on average, the global degree of common neighbors of nodes which ended up forming an edge is lower than the ones that did not. Thus, our findings on the Facebook dataset are consistent with those of  \citet{adamic2003friends} both before and after the introduction of PYMK. 

The behavior of personalized degree is shown in Figure \ref{fig:empirical-fb-mean-local}. Notice that personalized degree behaves in the exact \emph{opposite} manner compared to global degree. That is, common neighbors of the nodes which formed an edge with the ego node tend to have a \emph{higher} personalized degree, and this is true both before and after PYMK.

\subsection{Observations on Directed Networks} \label{sec:empirical-directed}

\begin{figure}[t]
    	\centering
    \includegraphics[width=2.8 in]{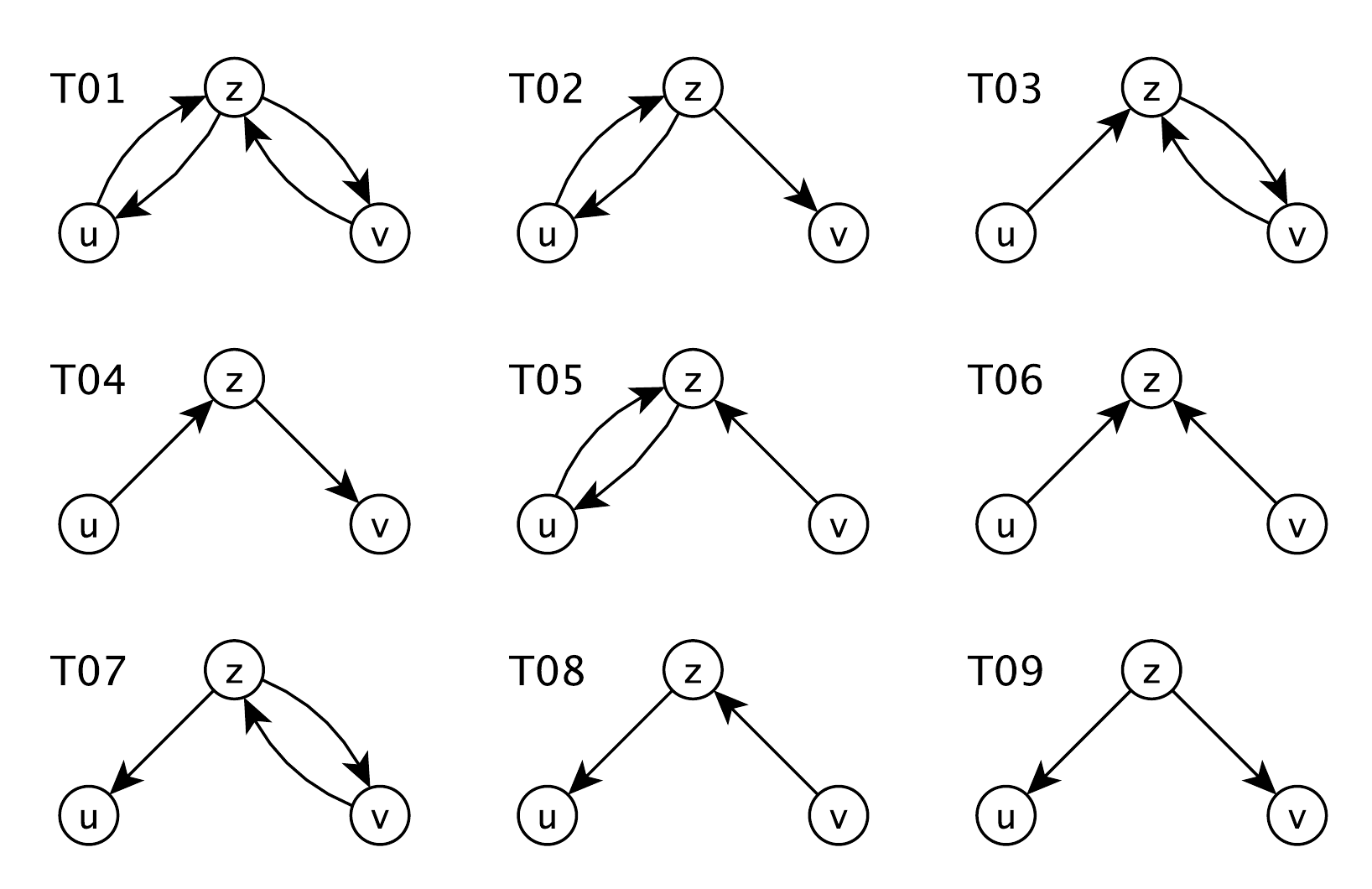}
    \caption{Different open triad patterns in a directed network. Triad labels match those used in \cite{schall2015social}.}
    \label{fig:triangleTypes}
\end{figure}

In the directed setting, we need to define which nodes we consider as an ego's neighbors and which nodes are 2 hops away. In Figure \ref{fig:triangleTypes}, consider node $u$ to be an ego, node $z$ to be ego's neighbor, and $v$ be a node that is two hops away from the ego. As depicted, there are 9 distinct possible ways that we can select nodes for our empirical analysis and that is without considering whether node $v$ follows the ego (node $u$).
In order to isolate the results for each triad type, we analyze each one separately. Additionally, we also need to consider the direction of edges. The rest of the empirical analysis will be the same as Section \ref{sec:empirical-procedure}.

\begin{figure}[t]
  \centering
  \subfloat{
  \includegraphics[width=2.5 in]{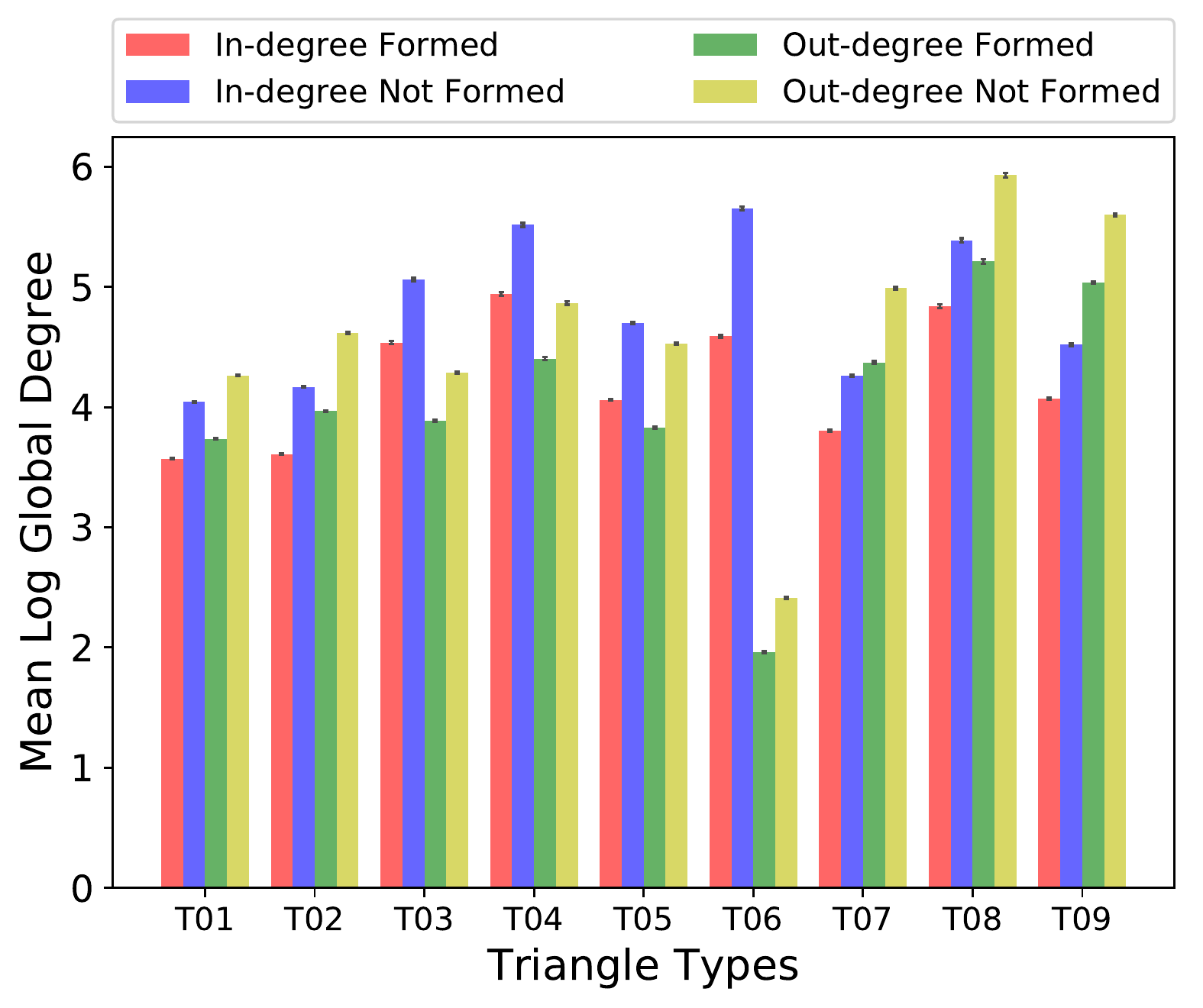}
  \label{fig:empirical-gplus-global-degree}}
  
  \subfloat{
  \includegraphics[width=2.5 in]{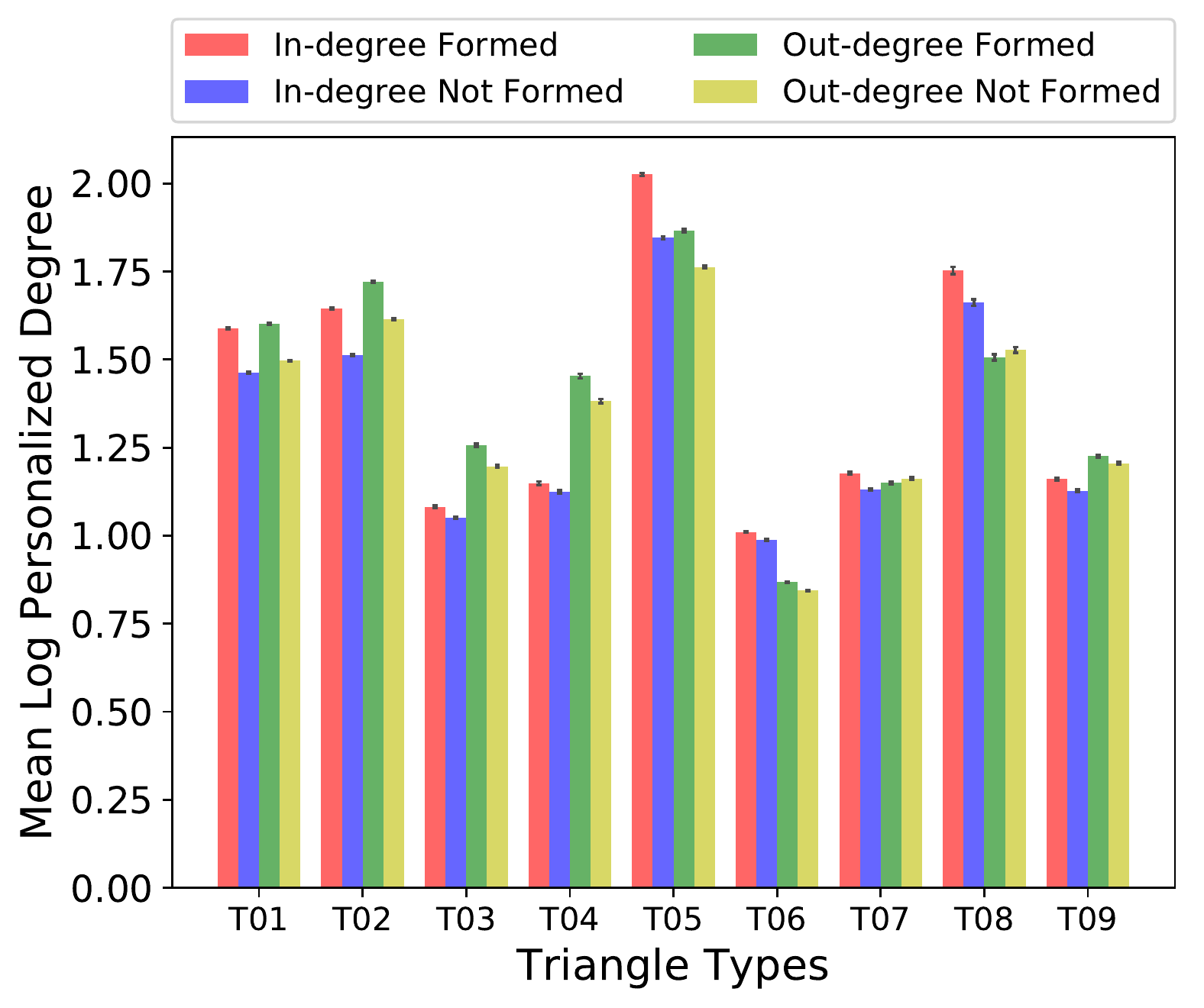}
  \label{fig:empirical-gplus-local-degree}}
  \caption{Mean log global and mean log personalized in- and out-degrees ($\pm$2 standard errors) of common neighbors in Google+ data (494,881 ego nodes).}
  \label{fig:empirical-gplus}
\end{figure}

\begin{figure}[t]
  \centering
  \subfloat[Flickr]{
  \includegraphics[width=2.5 in]{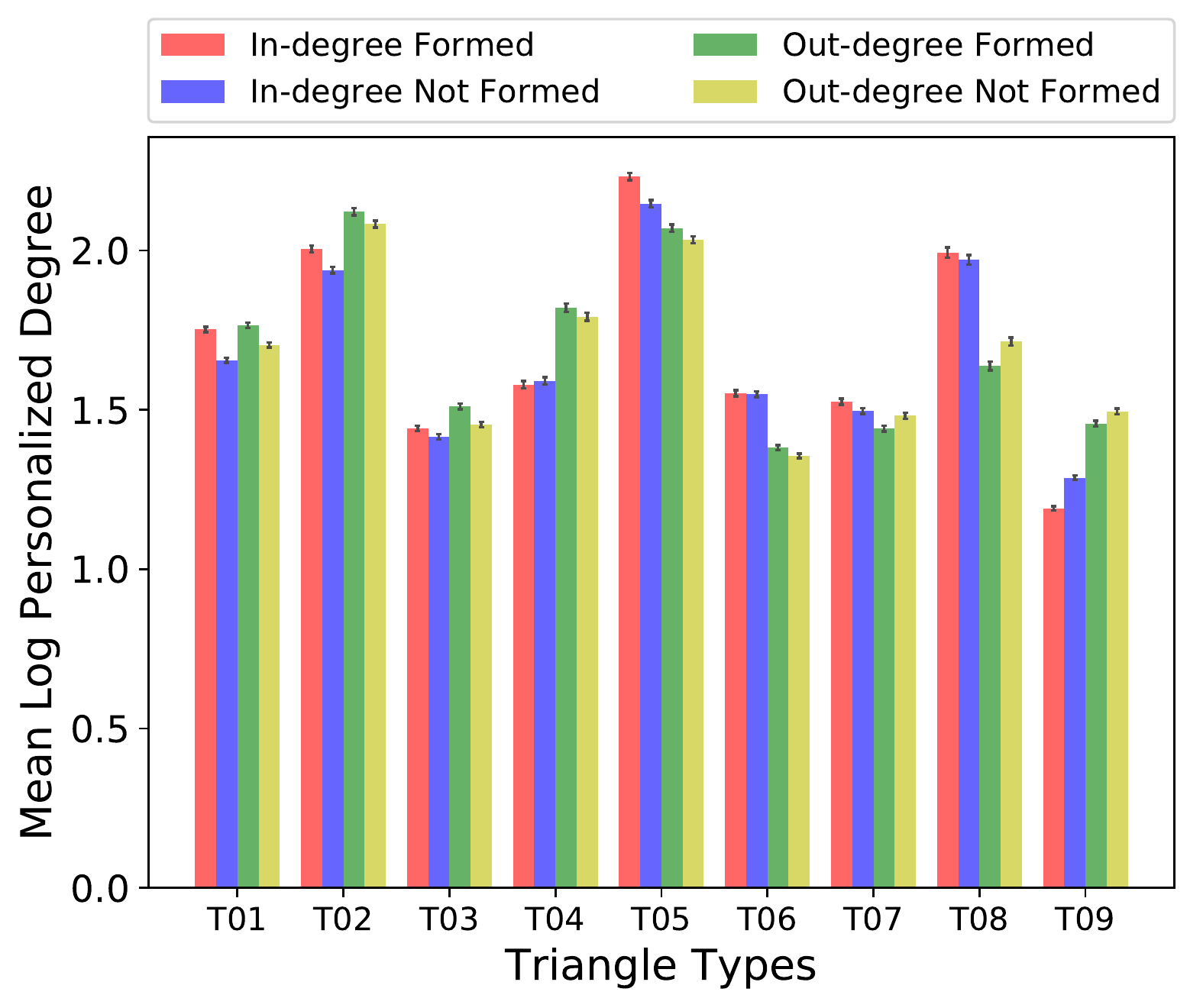}
  \label{fig:empirical-flickr-local-degree}}
  
  \subfloat[Digg]{
  \includegraphics[width=2.5 in]{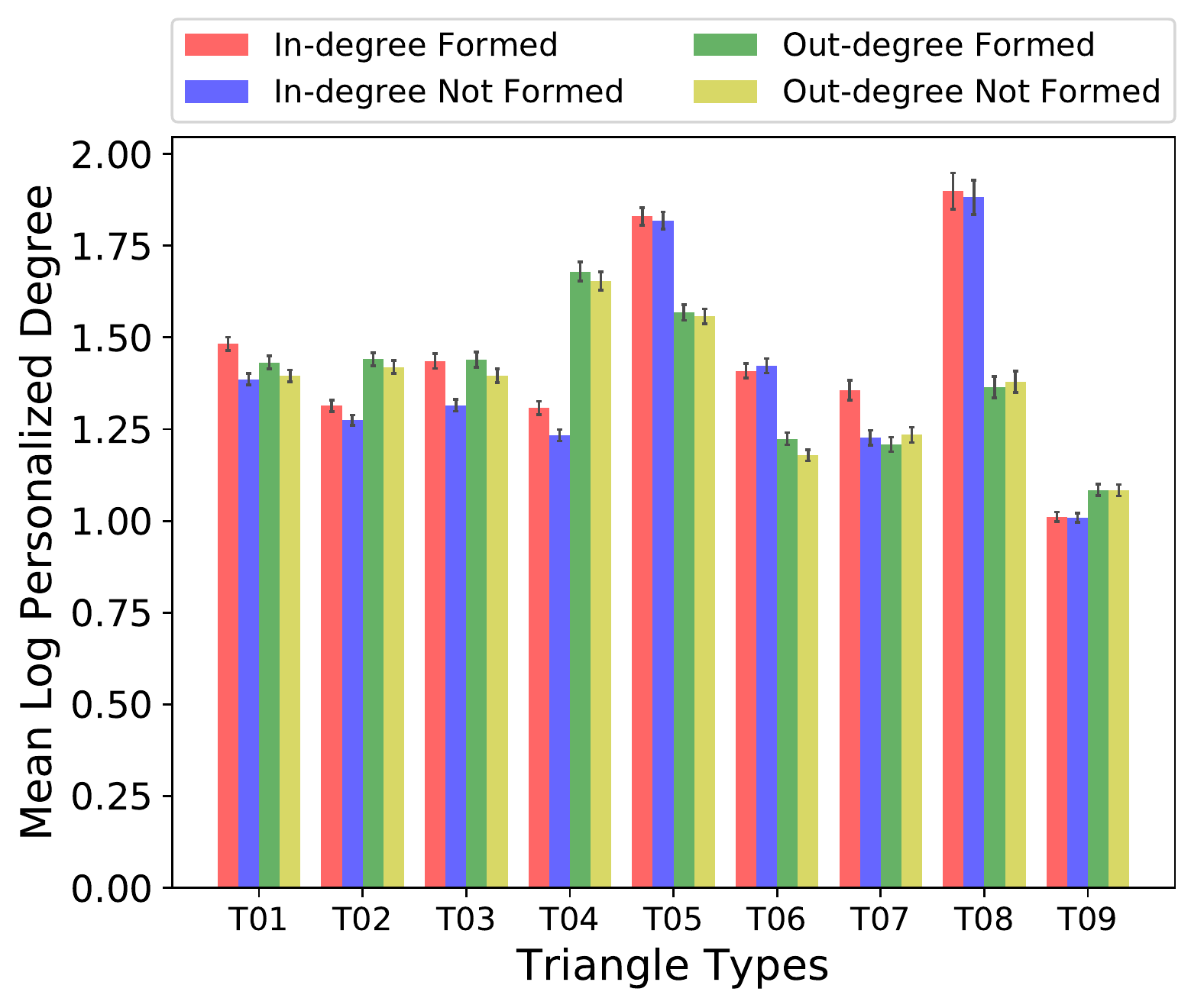}
  \label{fig:empirical-digg-local-degree}}
  \caption{Mean log personalized in- and out-degrees ($\pm$2 standard errors) of common neighbors in Flickr data (515,624 ego nodes) and Digg data (44,951 ego nodes).}
  \label{fig:empirical-flickr-digg}
\end{figure}

In directed settings our main findings are similar to those on the Facebook network. As shown in Figure \ref{fig:empirical-gplus-global-degree}, we observe that common neighbors with lower global in-degree or global out-degree are more predictive of future interactions in all 9 different triad types on Google+. This behaviour is also observed in both Flickr and Digg datasets. It is also conveyed from the figure that this observation holds much stronger for some triads than others (e.g. T01 vs. T06). Thereby, we can conclude that the findings of \citet{adamic2003friends} also hold in directed networks.

Similar to our findings in Section \ref{sec:empirical-facebook}, in the majority of triads in all directed datasets, personalized in- and out-degrees behave in an opposite manner compared to global degrees, as shown in Figure \ref{fig:empirical-gplus-local-degree} for Google+ and Figure \ref{fig:empirical-flickr-digg} for Flickr and Digg. That is, common neighbors with higher personalized in- and out-degrees are more predictive of future links. 
This holds much stronger for Google+, compared to Digg and Flickr, however the conclusion still holds in the majority of triad types on all three directed data sets. 
The larger standard errors in Figure \ref{fig:empirical-digg-local-degree} are due the smaller number of analyzed nodes in the Digg dataset.
The 3 triad types where it seems that this observation does not strongly hold  across networks are T07, T08, and T09. In all three cases, neighbors of the ego are considered to be ego's predecessors only. It is uncommon in most link prediction analyses to only consider a node's predecessors as its neighbors.

The main implication of this empirical analysis is that the high degree of a neighbor does not necessary make it less predictive of future links as implied by \citet{adamic2003friends}. If the high global degree of a common neighbor is caused by a high personalized degree, then that neighbor is in fact \emph{more predictive} of future links.

\section{Link Recommendation} \label{sec:link-recommendation}
Given our egocentric perspective approach concerning link formation, it is only fitting that we try to further validate our findings on personalized degree by incorporating it into link recommendation algorithms. 
If personalized degree does indeed play a crucial role in link formation, then a link recommendation algorithm that utilizes personalized degree should be more accurate than one that does not.

We first demonstrate a few possible ways to integrate our findings on personalized degree into simple neighborhood-based link recommendation algorithms, namely common neighbors \cite{liben2007link} and Adamic/Adar \citep{adamic2003friends}. Next, we evaluate these personalized degree-based methods on all four datasets. It is important to note that there are many more complex (and likely more accurate) link recommendation algorithms present in the literature as discussed in Section \ref{sec:related}. The link recommendation methods introduced in this section are quite simple and may not be the most accurate. However, the main purpose of this section is to first validate our empirical findings and consequently demonstrate the possibility for improving node neighborhood-based link recommendation algorithms by taking advantage of personalized degree. We believe that it is possible for more complex algorithms to benefit from personalized degree in a similar manner.

\subsection{Proposed Link Recommendation Methods}

\subsubsection{Personalized Degree Common Neighbors (PD-CN)} \label{sec:link-recommendation-ldcn}
We first integrate our empirical findings into the common neighbors link prediction method \citep{liben2007link}. The common neighbors (CN) link predictor assigns the score defined in \eqref{eq:cn-score} to each pair of nodes without a link. 

As we observed in Section \ref{sec:empirical}, neighbor nodes of the ego with a higher personalized degree are more predictive of future links. Thus, we should give these neighbor nodes higher weight in the link prediction score. We propose to sum over the log of personalized degrees of common neighbors. We take the log due to the heavy-tailed distribution of personalized degree as discussed in Section \ref{sec:personalized-degree-dist}.
Hence, the weight of each common neighbor will be the log of its personalized degree with the ego. Weighting common neighbors by degree has been used with global degree in Adamic/Adar, where common neighbors are weighted by the inverse log of their global degree \citep{adamic2003friends}. 

Therefore, PD-CN for an ego node $u$ and a candidate node $v$ is as follows:
\begin{equation} \label{eq:local-degree-common-neighbors}
	\PDCN(u, v) = \sum_{z\in \Gamma(u) \cap \Gamma(v)} \log{(|\Gamma (u) \cap \Gamma(z)| + 2)}
\end{equation}
where node $z$ is a common neighbor of nodes $u$ and $v$. 
The overlap of the neighborhoods of the ego node and the common neighbor correspond to the personalized degree of the common neighbor.
We shift personalized degree up by 2, since if it is 0 the log is undefined and if it is 1 then the log is 0, in which case the common neighbor is discarded from the sum.

\subsubsection{Personalized Degree Adamic/Adar (PD-AA)} \label{sec:link-recommendation-ldaa}
Here, we integrate personalized degree into the Adamic/Adar (AA) link predictor. 
Based on our empirical observations, both higher personalized degree and lower global degree of a neighbor node will lead to future link formation; therefore, the ratio of the personalized and global degrees can be a useful measure.

For every common neighbor node $z$, let $P_z = |\Gamma(u) \cap \Gamma(z)| + 1$ denote its personalized degree shifted up by 1 and $G_z = |\Gamma(z)| + 1$ denote its global degree shifted up by 1. Then, our proposed PD-AA score is given by
\begin{equation} \label{eq:local-degree-adamic-adar}
	\PDAA(u, v) 
    = \sum_{z\in \Gamma(u) \cap \Gamma(v)}{\left(\log\left[P_z\frac{(G_z - P_z)}{G_z} + G_z \frac{(G_z - P_z)}{P_z}\right]\right)^{-1}}
\end{equation}
PD-AA score tries to keep the intuition behind AA, that a higher global degree should result in a lower score while also allowing a higher personalized degree to result in a higher score as well. In order to avoid division by zero, we shift personalized degree up by 1 and since it is a subset of global degree we similarly need to shift up the global degree.

Notice that the overall structure of \eqref{eq:local-degree-adamic-adar} is very similar to the AA score \eqref{eq:aa-score}. In order to achieve a good balance between global and personalized degrees, we separate the term inside the log into two parts: the weight with respect to $P_z$ and the weight with respect to  $G_z$. We also use a ratio involving the difference between ($G_z - P_z$).
Here, as $P_z$ approaches 1, PD-AA becomes a harsher version of AA with almost 2 times the penalization for global degree, and as $P_z$ approaches $G_z$, it tries to subtract twice the personalized degree from global degree to decrease the penalty for higher global degree.

\subsection{Evaluation Approach} \label{sec:link-recommendation-eval}
We test both of our proposed methods, PD-CN and PD-AA, and compare them to CN and AA, respectively. For every snapshot of a network, we compute each of the link prediction scores for all node pairs $(u, v)$ where $u$ is an ego node and $v$ is a node that is 2 hops away from the ego. Next, for each method, we rank all nodes 2 hops away from the ego based on the method's score in deceasing order. Then, we choose top-K predictive rate, also known as precision at K (P@K), to be our evaluation metric. It is the percentage of correctly classified positive samples among the top K instances in the ranking by a specified link predictor. It has been used by others for link recommendation \cite{Yin2010,Backstrom2010}, and we view it as the most relevant accuracy metric in a recommendation setting since these are the people who are potentially going to be recommended to a user by a feature such as PYMK in Facebook. Since P@K is heavily dependent on K \citep{Yang2015}, we chose several K values in all of our evaluations. For each method used in an egocentric network, its top-K predictive rate at a specific K is the mean over all snapshots.

\subsection{Results on Facebook Data} \label{sec:link-recommendation-fb}

\begin{figure}[t]
  \centering
  \subfloat[Before PYMK]{
  \includegraphics[width=0.48\columnwidth]{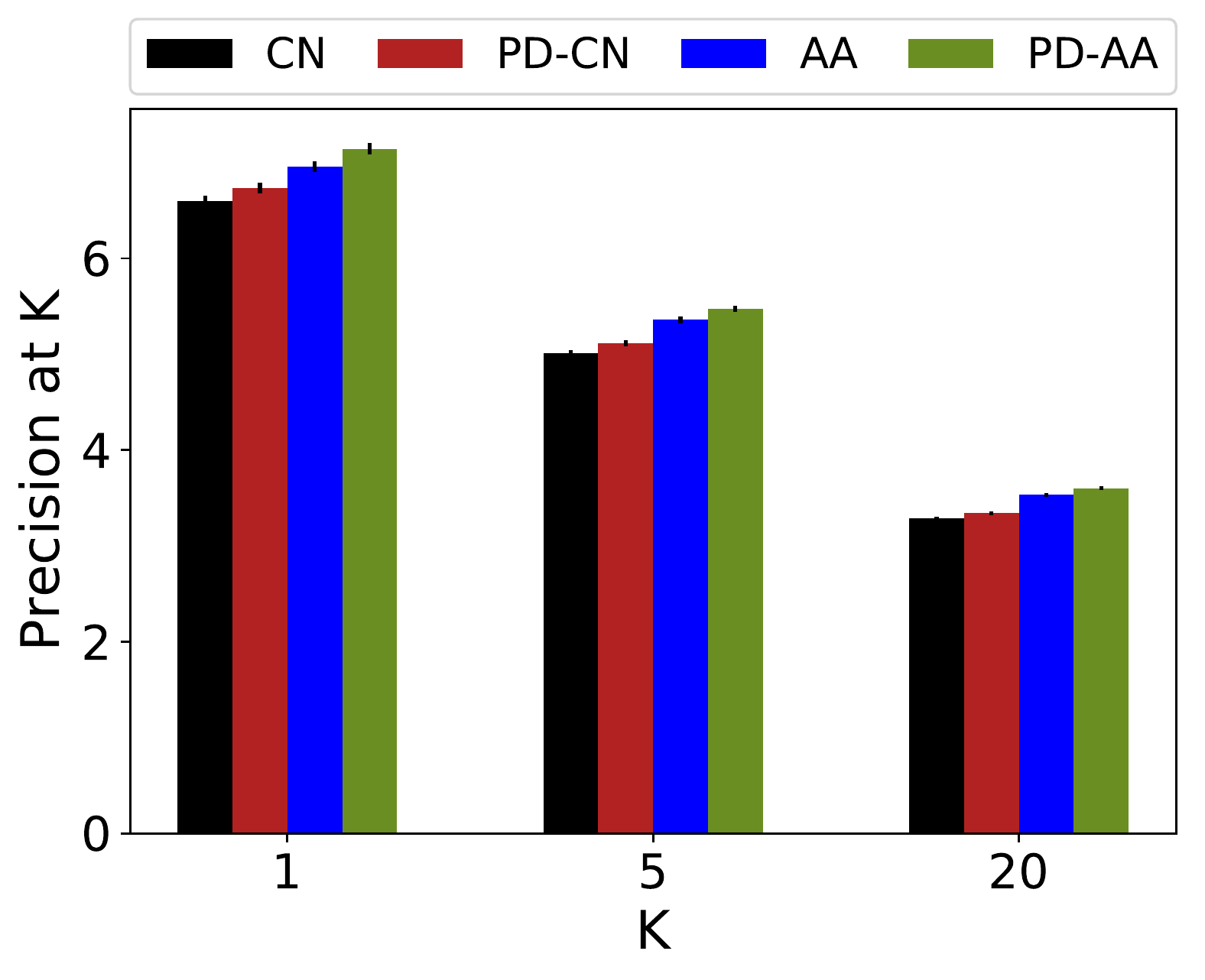}
  \label{fig:fb-lp-performance-before-pymk}}
  \subfloat[After PYMK]{
  \includegraphics[width=0.48\columnwidth]{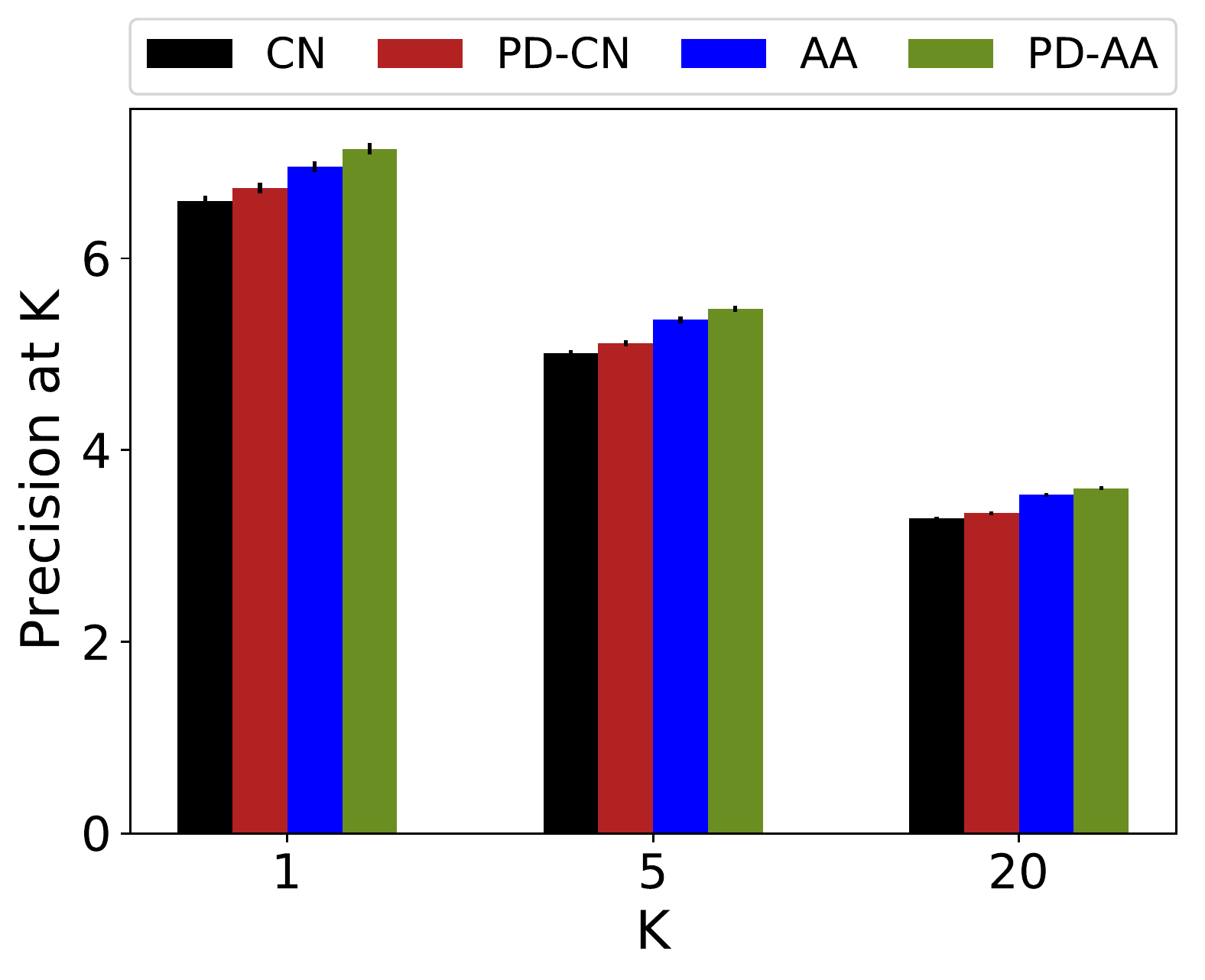}
  \label{fig:fb-lp-performance-after-pymk}}
  \caption{Mean top-K predictive rate of all methods ($\pm$ standard error) for the Facebook data.}
  \label{fig:fb-lp-performance}
\end{figure}

\begin{figure}[t]
  \centering
  \subfloat[Before PYMK]{
  \includegraphics[width=0.48\columnwidth]{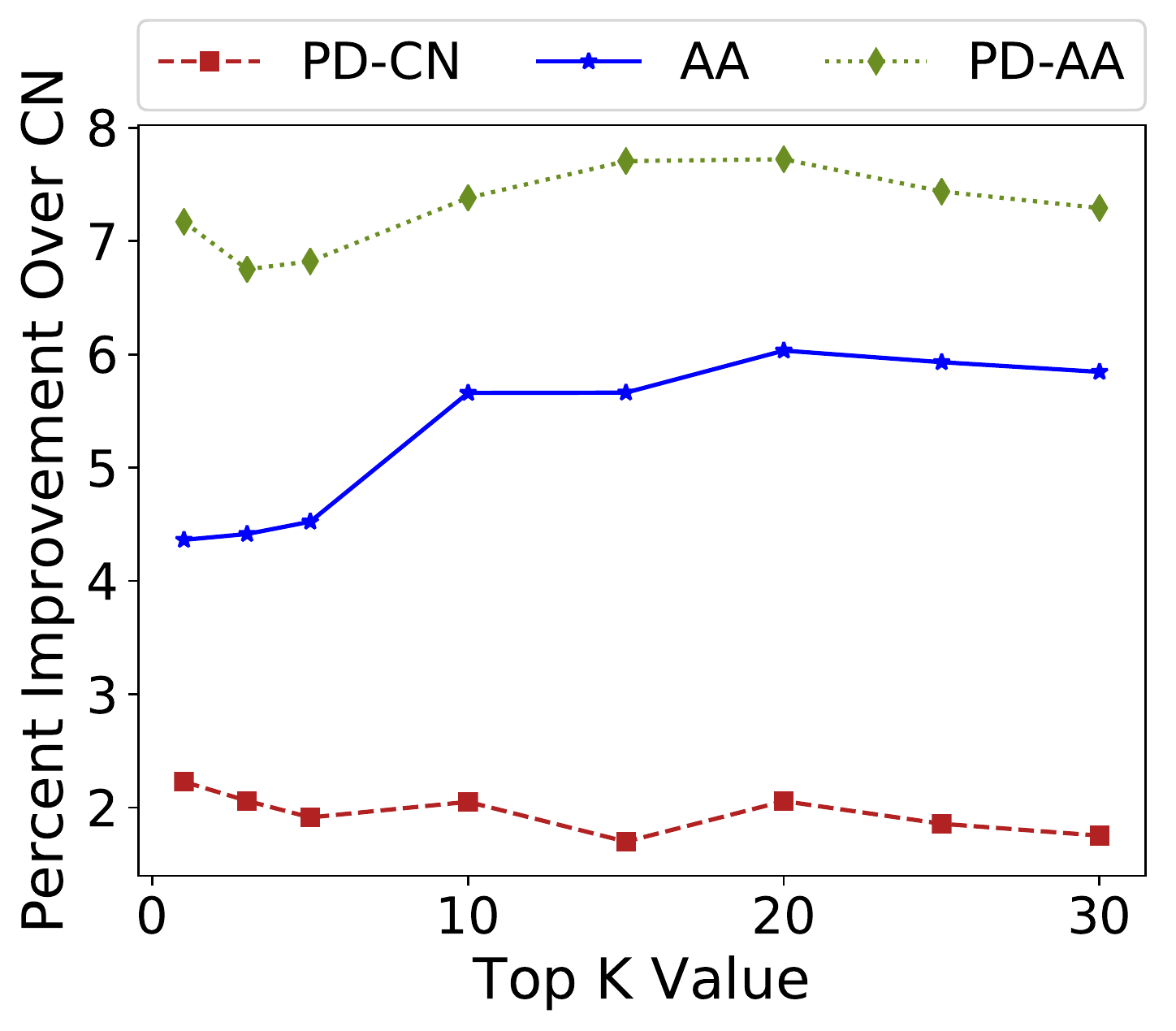}
  \label{fig:fb-lp-percent-imp-before-pymk}}
  \subfloat[After PYMK]{
  \includegraphics[width=0.48\columnwidth]{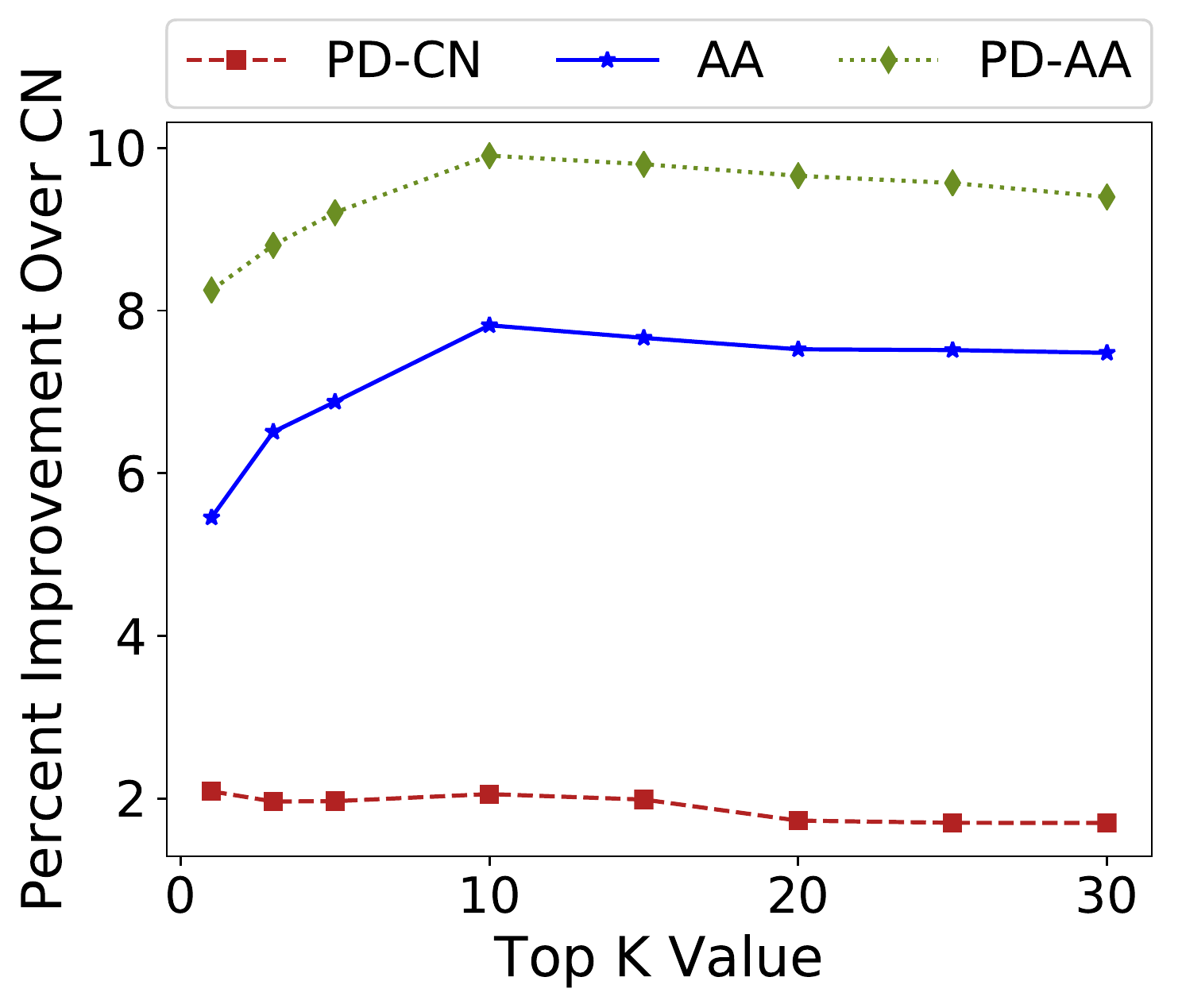}
  \label{fig:fb-lp-percent-imp-after-pymk}}
  \caption{Percent improvements of all link prediction metrics compared to common neighbors as the base metric for Facebook data.}
  \label{fig:fb-lp-percent-imp}
\end{figure}

The link recommendation results for each of the algorithms on the Facebook dataset, both before and after the introduction of PYMK are shown in Figure \ref{fig:fb-lp-performance}. 
Notice that, for each value of K, incorporating personalized degree (PD-CN) consistently improves over CN. 
However, incorporating global degree (AA) has a greater improvement on link recommendation accuracy.
The greatest improvement is seen when both personalized and global degrees are included (PD-AA).
We also calculate the percentage improvement of all measures with respect to common neighbors, as shown in Figure \ref{fig:fb-lp-percent-imp}. Here we can also observe the improvements in each method caused by the incorporation of personalized degree. 
We find that incorporating personalized degrees improves link recommendation accuracy by about $2\%$, incorporating global degree (AA) improves it by about $4\%$ to $8\%$, and incorporating both improves it by about $6\%$ to $10\%$. 
These improvements further validate our observations regarding the behavior of personalized degree.

\subsection{Results on Directed Networks} \label{sec:link-recommendation-gplus}

\begin{figure}[t]
  \centering
  \subfloat[Google+]{
  \includegraphics[width=0.47\columnwidth]{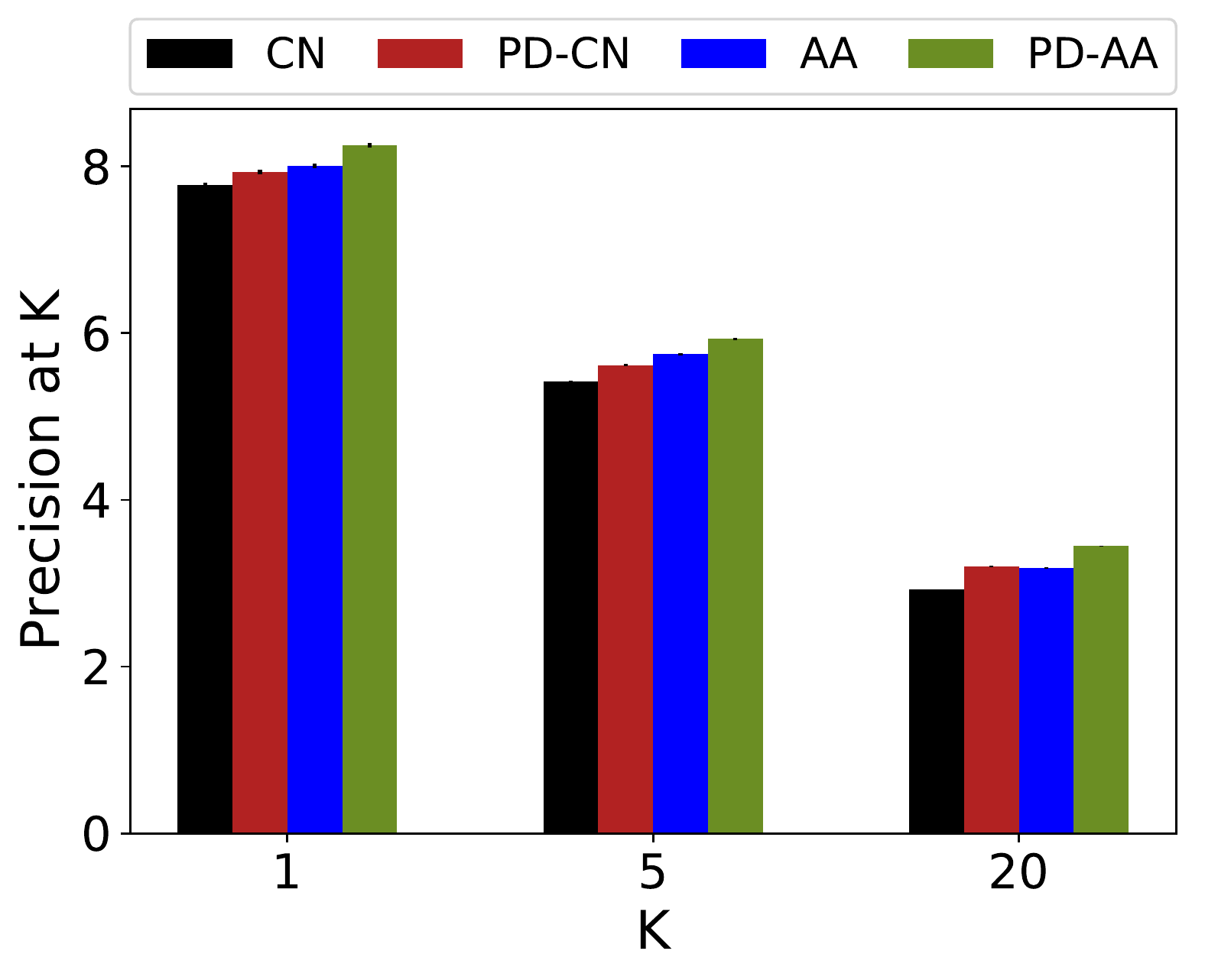}
  \label{fig:gplus-lp-performance}}
  \subfloat[Flickr]{
  \includegraphics[width=0.47\columnwidth]{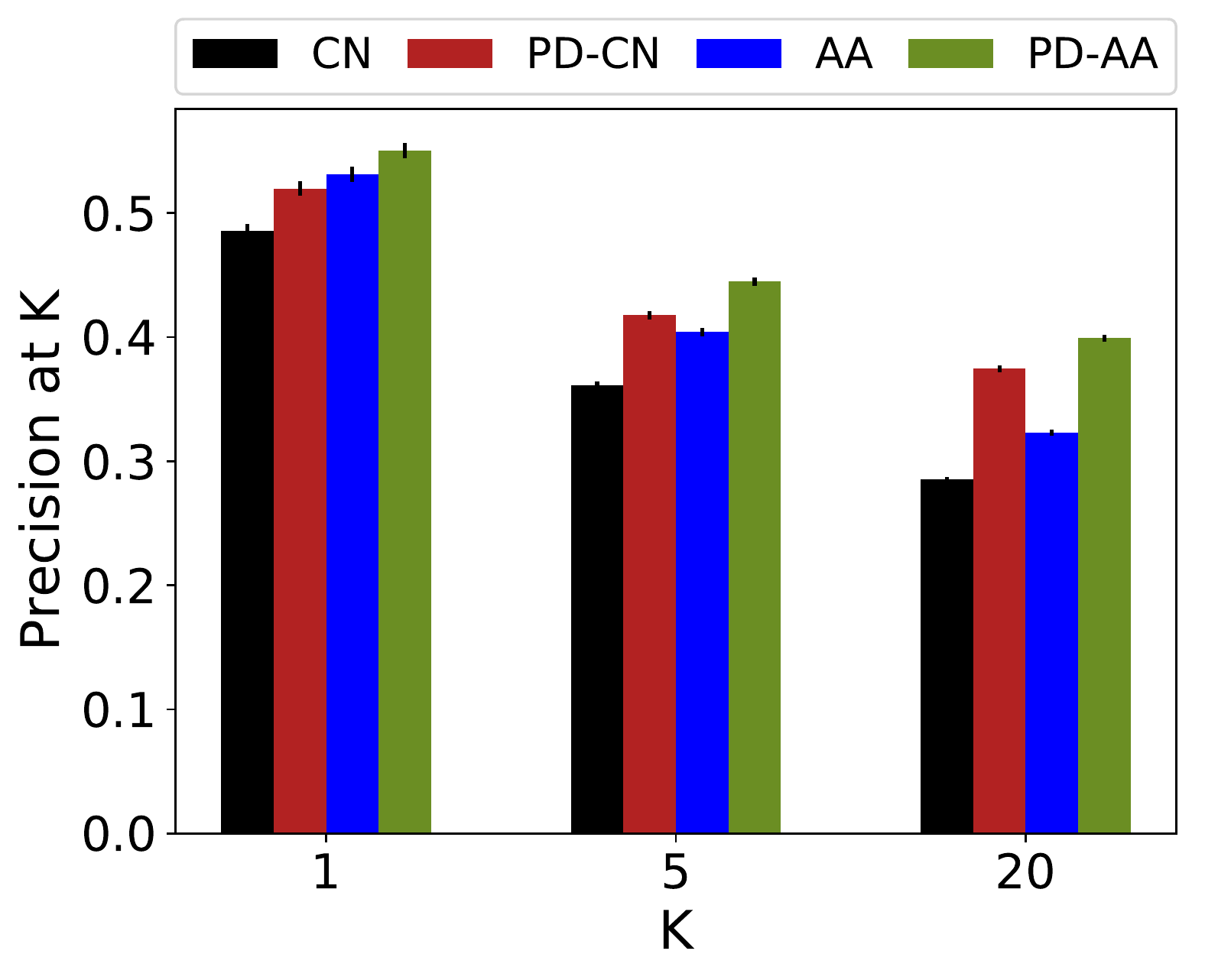}
  \label{fig:flickr-lp-performance}}
  
  \subfloat[Digg]{
  \includegraphics[width=0.475\columnwidth]{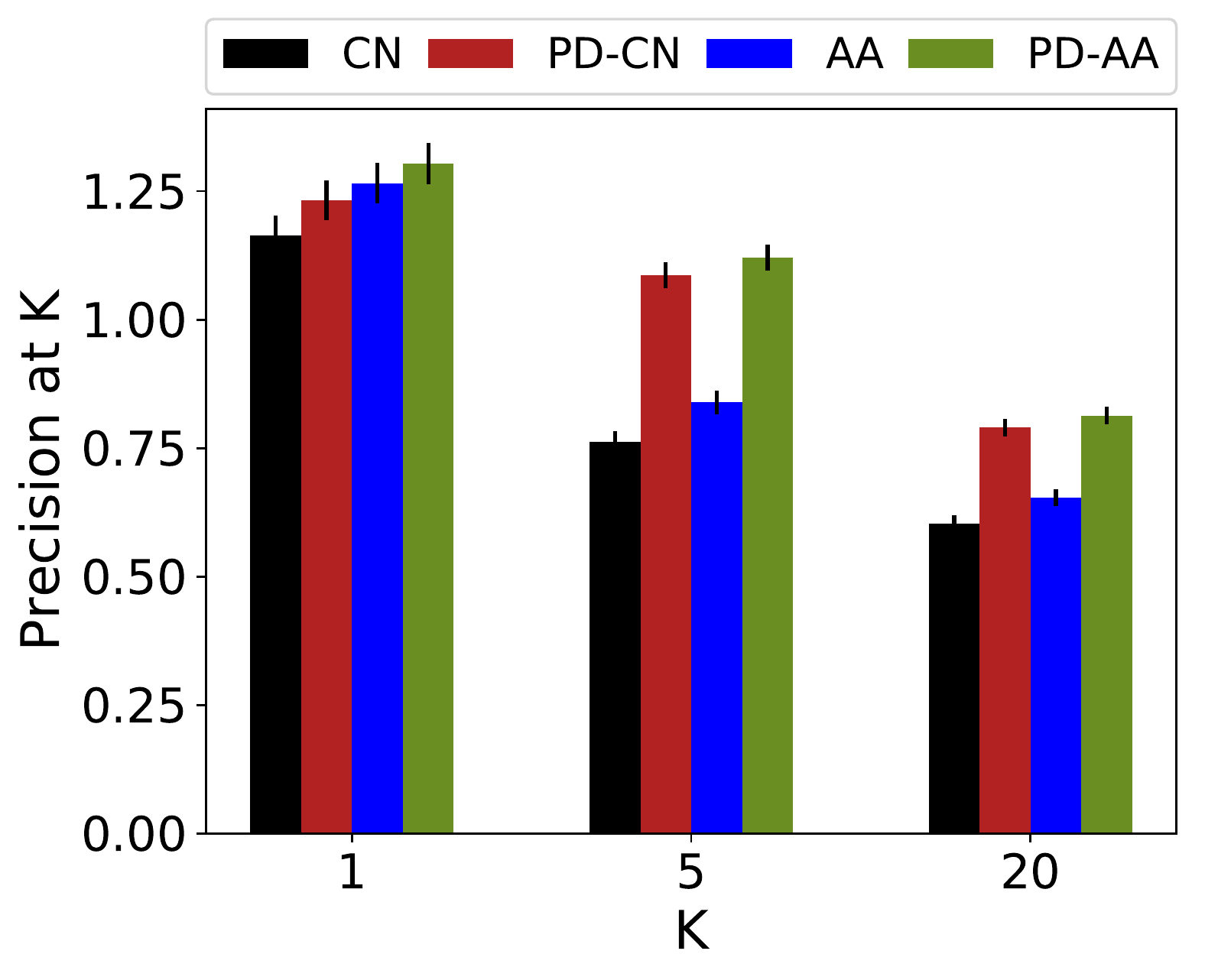}
  \label{fig:digg-lp-performance}}
  
  \caption{Mean top-K predictive rate of all (undirected) methods ($\pm$ standard error) for directed datasets.}
  \label{fig:directed-lp-performance}
\end{figure}

\begin{figure}[t]
  \centering
  \subfloat[Google+]{
  \includegraphics[width=0.48\columnwidth]{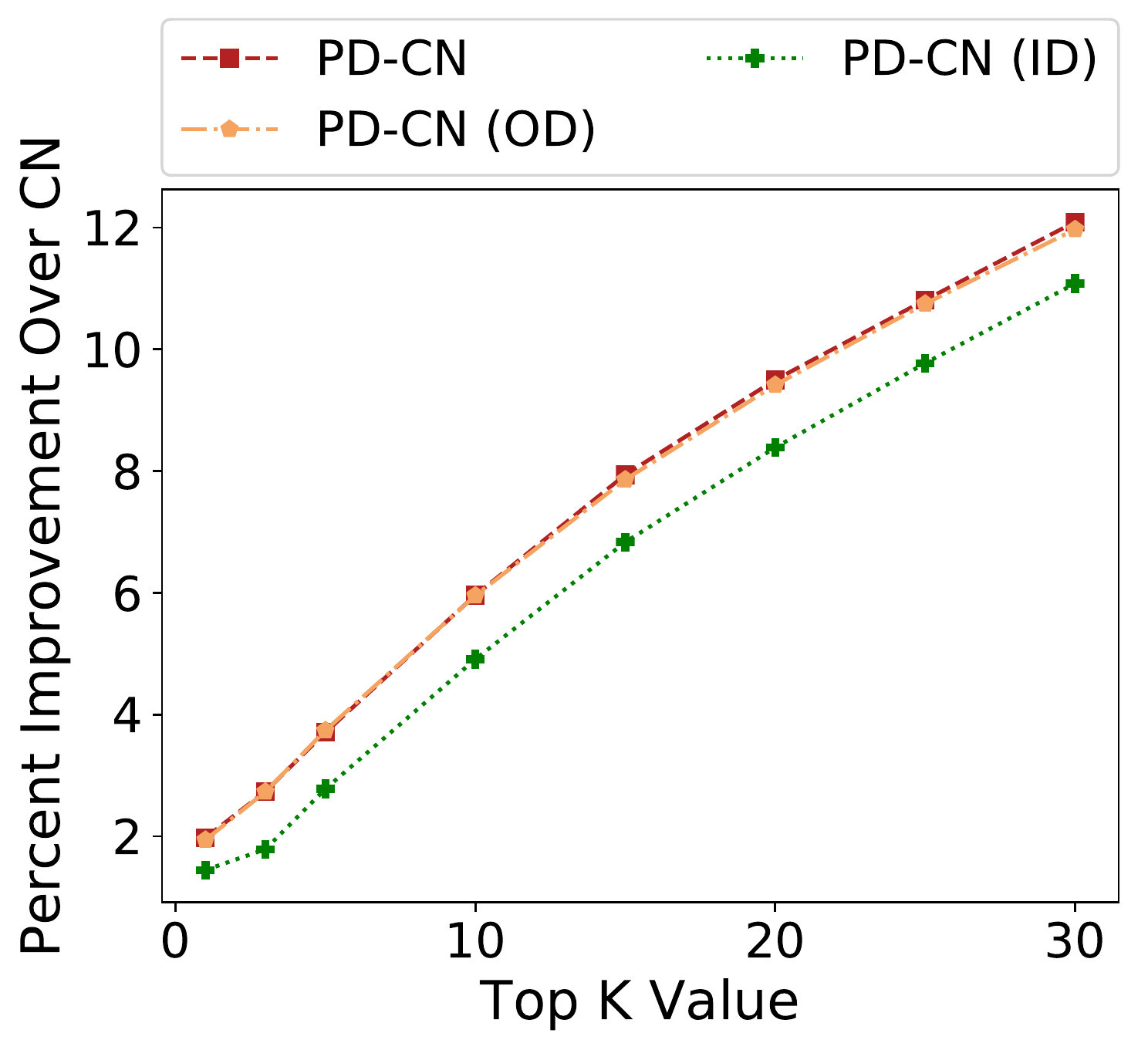}
  \includegraphics[width=0.48\columnwidth]{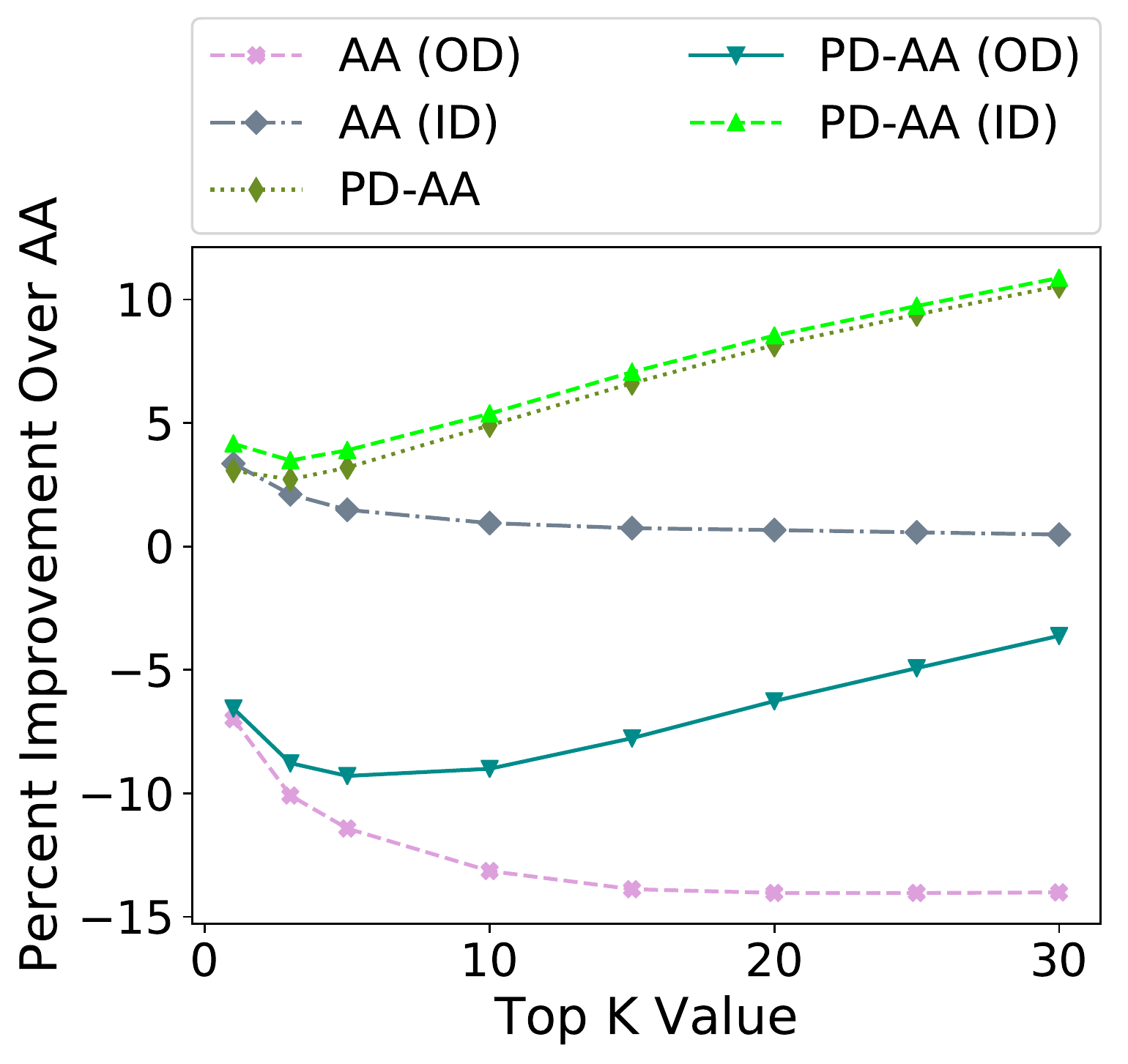}
  \label{fig:gplus-lp-percent-imp}}
  
  \subfloat[Flickr]{
  \includegraphics[width=0.48\columnwidth]{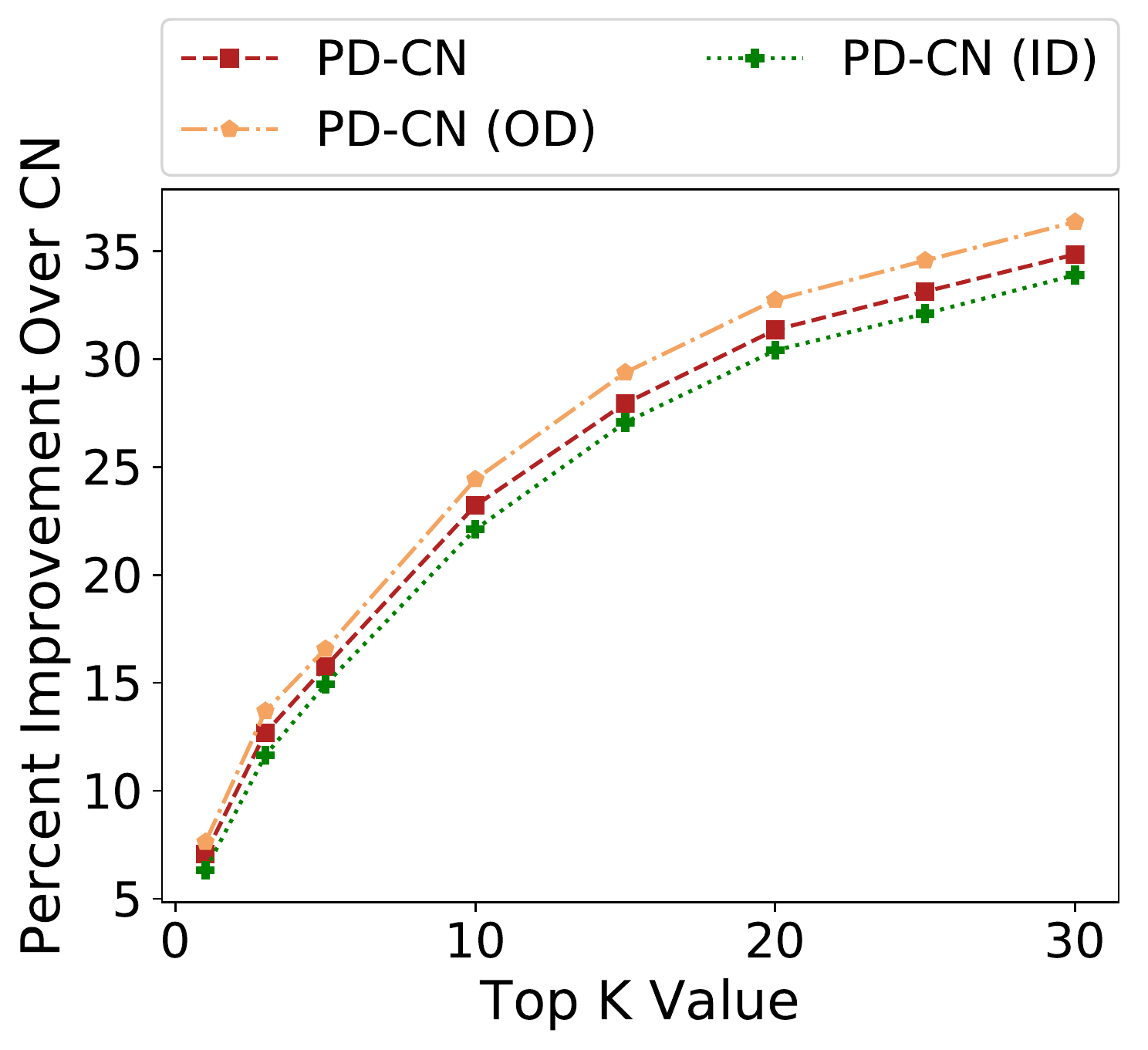}
  \includegraphics[width=0.48\columnwidth]{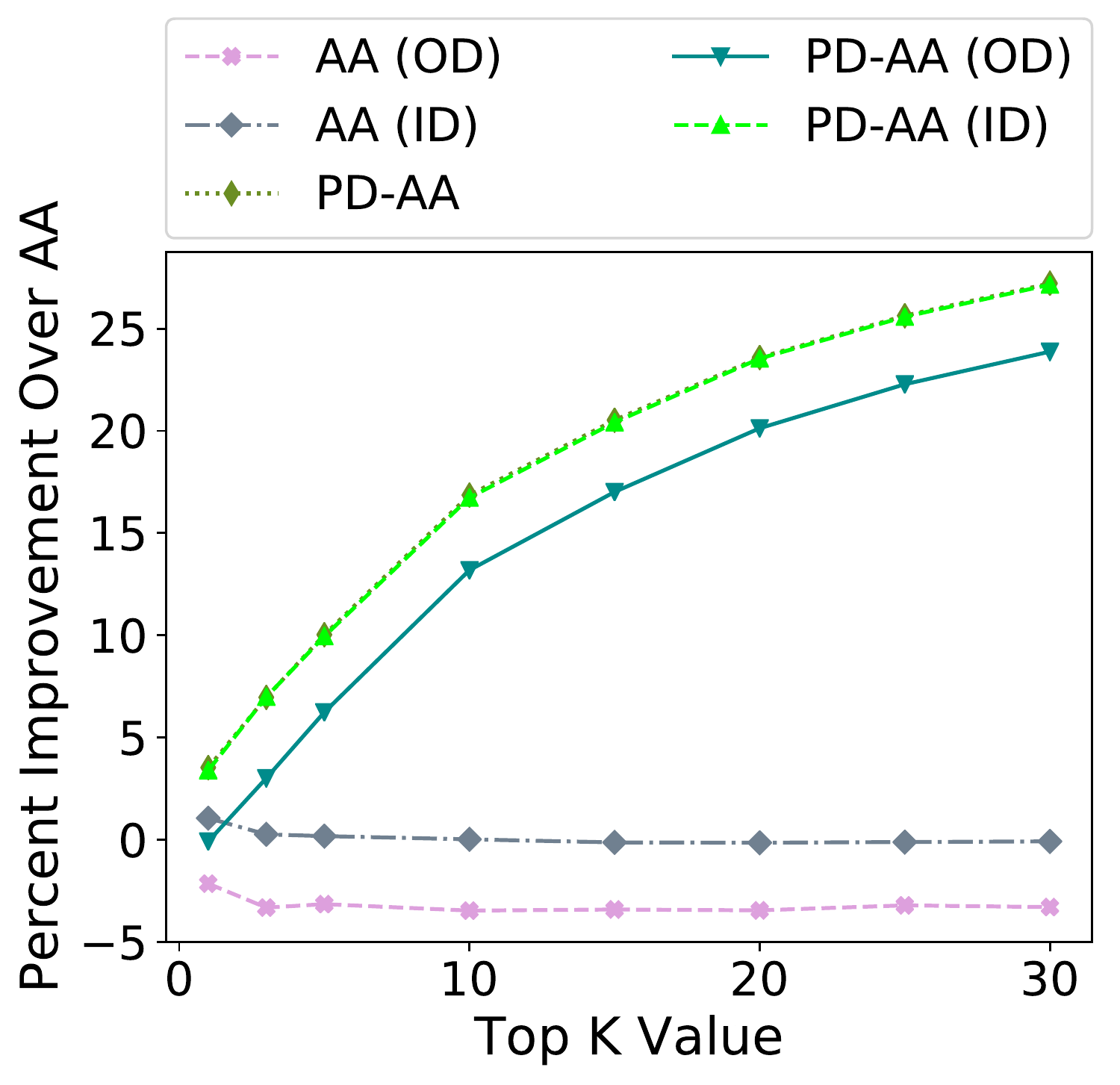}
  \label{fig:flickr-lp-percent-imp}}
  
  \subfloat[Digg]{
  \includegraphics[width=0.48\columnwidth]{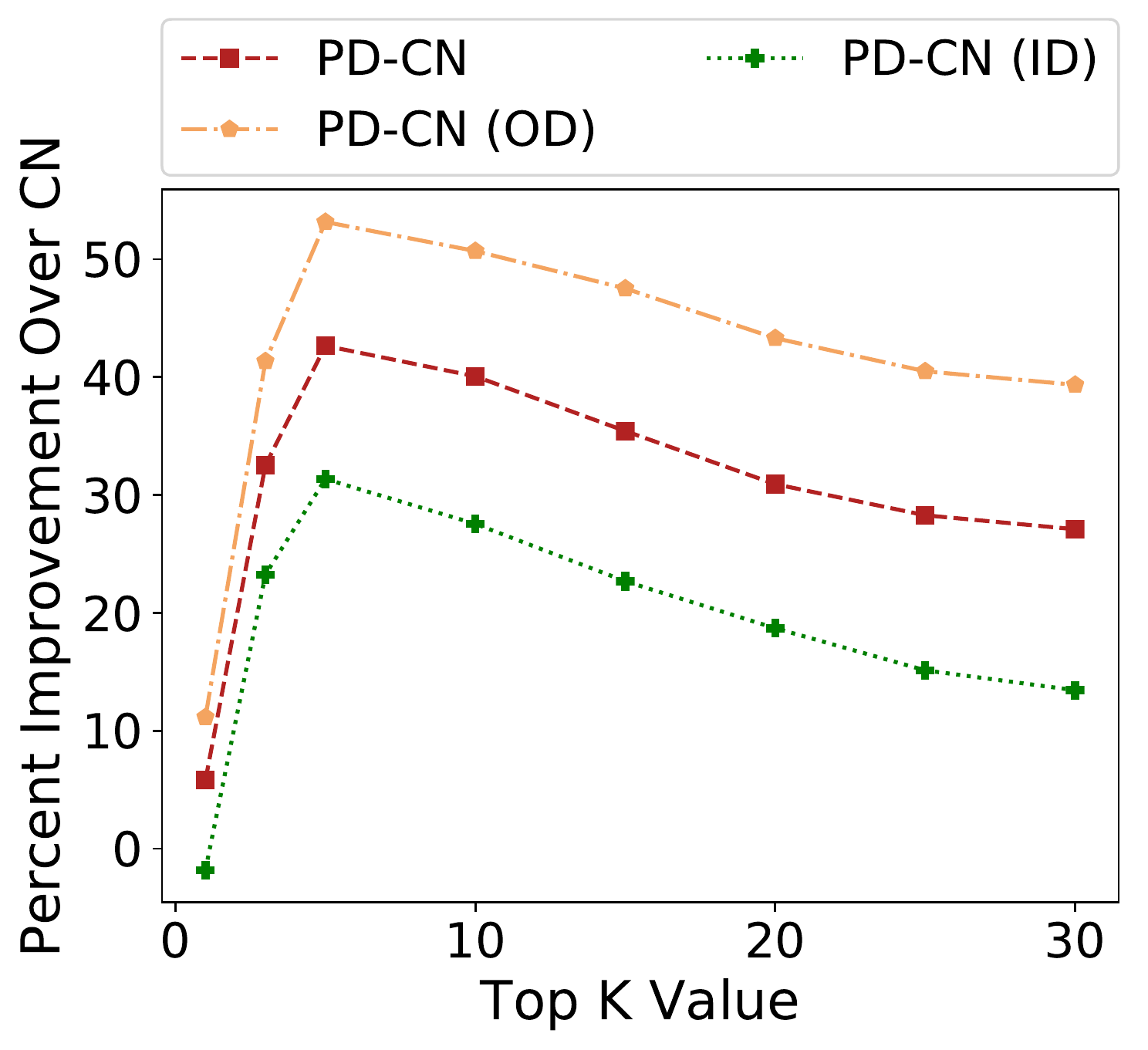}
  \includegraphics[width=0.48\columnwidth]{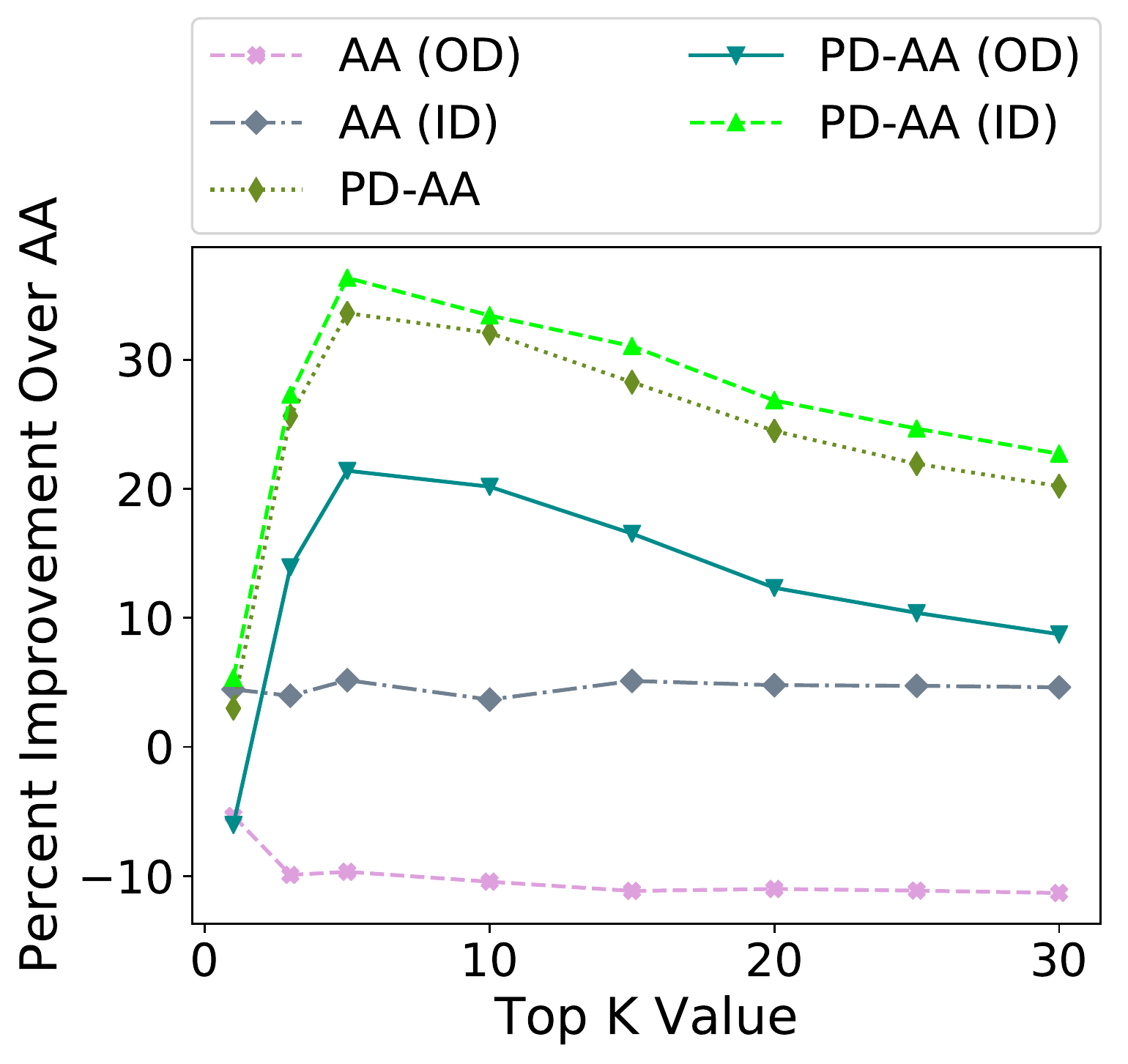}
  \label{fig:digg-lp-percent-imp}}
  
  \caption{Percent improvements of all CN- and AA-based in- and out-degree link prediction metrics, respectively compared to CN and AA as the base metric. \textit{(ID)} and \textit{(OD)} indicate the \textit{in-degree only} and \textit{out-degree only} variants, respectively.}
  \label{fig:directed-lp-percent-imp}
\end{figure}

Similar to our prior directed analyses, we consider all successors of an ego node as its neighbors ($z$ nodes), and all the successors and predecessors of each neighbor node, that are not themselves neighbors of the ego, to be the nodes that are two hops away form the ego ($v$ nodes).
Additionally, given that in a directed setting we also have access to the direction of edges, we consider three variants of each method, except for CN since the degree of a node is not a factor.
The first variant is turning the network into undirected by reciprocating all edges, the second one is only using in-degree, and lastly only using out-degree. In the case of in- and out-degree variants for AA, we shift up both in- and out-degrees by 2 to avoid taking the log of or dividing by 0. Similarly, in case of PD-AA, since there is no guarantee for global in- or out-degree to be higher than personalized in- and out- degree, respectively, we add 1 to the definition of $G_z$ as described in \ref{eq:local-degree-adamic-adar}. The evaluation approach is the same as explained in Section \ref{sec:link-recommendation-eval}.

The results for all three directed datasets are shown in Figures \ref{fig:directed-lp-performance} and \ref{fig:directed-lp-percent-imp}. 
We note that link formation in both Flickr and Digg occurs a lot less frequently compared to Google+, and thus, it becomes a much harder task to accurately predict future links, resulting in the much lower values of P@K for Flickr and Digg in Figure \ref{fig:directed-lp-performance}, but trends are consistent across datasets. 
In Google+, Flickr, and Digg, PD-CN improved CN by an average of $12\%$, $23\%$ and $30\%$, respectively (averaged over values of K). Moreover, only considering global degree (AA) improved prediction accuracy over CN, as well, by $9.4\%$, $12\%$, and $9.0\%$ for Google+, Flickr, and Digg, respectively. However, unlike our results in the undirected setting, only considering personalized degree (PD-CN) results in much higher improvements compared to only considering global degree (AA). 

Additionally, as shown in Figure \ref{fig:directed-lp-percent-imp}, although all three variants of PD-CN will result in an improvement over CN, only considering personalized in-degree will decrease P@K compared to the undirected method, while only utilizing personalized out-degree may improve the overall accuracy, sometimes by as much as 10\%, as shown in Figure \ref{fig:digg-lp-percent-imp}. 
Similarly, analyzing the effects of in- and out-degree on AA, it is clear that limiting global degree to out-degree will result in a decrease in P@K for all K values, while limiting it to in-degree may result in improvements as it is depicted in Figure \ref{fig:digg-lp-percent-imp}. Moreover, while all three variants of PD-AA out-perform all variants of AA in all datasets for all values of K, the impacts of degree directionality on PD-AA are similar to that of AA. Thus, if we consider global degree, personalized degree, as well as the direction of edges, the in-degree variant of PD-AA outperforms all other methods. Comparing it to CN as the base method, average percent improvements for Google+, Flickr, and Digg are $21\%$, $31\%$ and $37\%$, respectively.
Hence, it is safe to conclude that utilizing global in-degree and personalized out-degree can boost link recommendation accuracy.

\section{Conclusion}
In this paper, we examined the role of node degrees in link formation from an egocentric perspective. We introduced the notion of a personalized node degree for the neighbors of the ego as the number of other neighbors a particular neighbor is connected to. 
Based on the result of our empirical study on four different datasets, we conclude that the personalized degree of common neighbors behaves in the opposite manner of their global degree when it comes to link formation. While \citet{adamic2003friends} found (and we also verified on our datasets) that common neighbors with high global degree are less likely to be predictive of future links, we find that common neighbors with higher personalized degree are \emph{more likely} to be predictive of future links.
We also demonstrated an approach to incorporate our empirical findings into node neighborhood-based link recommendation algorithms, and we were able to verify that it leads to improvements in link recommendation accuracy. Furthermore, in the case of directed links, we found personalized out-degree and global in-degree to be more predictive than personalized in-degree and global out-degree, respectively.

We believe our findings have many implications towards understanding link formation from an egocentric perspective. 
We demonstrated how our findings can be used to improve link recommendation algorithms. 
Another potential application is towards generative models for network growth. Such models typically operate at a global level, and node degree plays an important role in many models, including the model of \citet{barabasi1999emergence} to generate scale-free networks. Incorporating personalized degree into such models may allow for more accurate replication of link formations in real networks.

\begin{acks}
This material is based upon work supported by the \grantsponsor{nsf}{National Science Foundation}{https://nsf.gov} grant \grantnum{nsf}{IIS-1755824}.
We would like to thank the University of Toledo's Office of Undergraduate Research for providing funding for Makan Arastuie through the Undergraduate Summer Research and Creative Activities Program (USRCAP).
\end{acks}

\bibliographystyle{ACM-Reference-Format}
\balance 
\bibliography{ms}


\begin{thebibliography}{22}


\ifx \showCODEN    \undefined \def \showCODEN     #1{\unskip}     \fi
\ifx \showDOI      \undefined \def \showDOI       #1{#1}\fi
\ifx \showISBNx    \undefined \def \showISBNx     #1{\unskip}     \fi
\ifx \showISBNxiii \undefined \def \showISBNxiii  #1{\unskip}     \fi
\ifx \showISSN     \undefined \def \showISSN      #1{\unskip}     \fi
\ifx \showLCCN     \undefined \def \showLCCN      #1{\unskip}     \fi
\ifx \shownote     \undefined \def \shownote      #1{#1}          \fi
\ifx \showarticletitle \undefined \def \showarticletitle #1{#1}   \fi
\ifx \showURL      \undefined \def \showURL       {\relax}        \fi
\providecommand\bibfield[2]{#2}
\providecommand\bibinfo[2]{#2}
\providecommand\natexlab[1]{#1}
\providecommand\showeprint[2][]{arXiv:#2}

\bibitem[\protect\citeauthoryear{Adamic and Adar}{Adamic and Adar}{2003}]%
        {adamic2003friends}
\bibfield{author}{\bibinfo{person}{Lada~A. Adamic} {and} \bibinfo{person}{Eytan
  Adar}.} \bibinfo{year}{2003}\natexlab{}.
\newblock \showarticletitle{Friends and neighbors on the web}.
\newblock \bibinfo{journal}{\emph{Social Networks}} \bibinfo{volume}{25},
  \bibinfo{number}{3} (\bibinfo{year}{2003}), \bibinfo{pages}{211--230}.
\newblock


\bibitem[\protect\citeauthoryear{Backstrom and Leskovec}{Backstrom and
  Leskovec}{2011}]%
        {Backstrom2010}
\bibfield{author}{\bibinfo{person}{Lars Backstrom} {and} \bibinfo{person}{Jure
  Leskovec}.} \bibinfo{year}{2011}\natexlab{}.
\newblock \showarticletitle{Supervised random walks: Predicting and
  recommending links in social networks}. In
  \bibinfo{booktitle}{\emph{Proceedings of the 4th ACM International Conference
  on Web Search and Data Mining}}. \bibinfo{pages}{635--644}.
\newblock


\bibitem[\protect\citeauthoryear{Barab{\'a}si and Albert}{Barab{\'a}si and
  Albert}{1999}]%
        {barabasi1999emergence}
\bibfield{author}{\bibinfo{person}{Albert-L{\'a}szl{\'o} Barab{\'a}si} {and}
  \bibinfo{person}{R{\'e}ka Albert}.} \bibinfo{year}{1999}\natexlab{}.
\newblock \showarticletitle{Emergence of scaling in random networks}.
\newblock \bibinfo{journal}{\emph{Science}} \bibinfo{volume}{286},
  \bibinfo{number}{5439} (\bibinfo{year}{1999}), \bibinfo{pages}{509--512}.
\newblock


\bibitem[\protect\citeauthoryear{Barab{\'a}si and Bonabeau}{Barab{\'a}si and
  Bonabeau}{2003}]%
        {barabasi2003scale}
\bibfield{author}{\bibinfo{person}{Albert-L{\'a}szl{\'o} Barab{\'a}si} {and}
  \bibinfo{person}{Eric Bonabeau}.} \bibinfo{year}{2003}\natexlab{}.
\newblock \showarticletitle{Scale-free networks}.
\newblock \bibinfo{journal}{\emph{Scientific American}} \bibinfo{volume}{288},
  \bibinfo{number}{5} (\bibinfo{year}{2003}), \bibinfo{pages}{60--69}.
\newblock


\bibitem[\protect\citeauthoryear{Cannistraci, Alanis-Lobato, and
  Ravasi}{Cannistraci et~al\mbox{.}}{2013}]%
        {cannistraci2013link}
\bibfield{author}{\bibinfo{person}{Carlo~Vittorio Cannistraci},
  \bibinfo{person}{Gregorio Alanis-Lobato}, {and} \bibinfo{person}{Timothy
  Ravasi}.} \bibinfo{year}{2013}\natexlab{}.
\newblock \showarticletitle{From link-prediction in brain connectomes and
  protein interactomes to the local-community-paradigm in complex networks}.
\newblock \bibinfo{journal}{\emph{Scientific Reports}}  \bibinfo{volume}{3}
  (\bibinfo{year}{2013}).
\newblock


\bibitem[\protect\citeauthoryear{Gong, Xu, Huang, Mittal, Stefanov, Sekar, and
  Song}{Gong et~al\mbox{.}}{2012}]%
        {gong2012evolution}
\bibfield{author}{\bibinfo{person}{Neil~Zhenqiang Gong},
  \bibinfo{person}{Wenchang Xu}, \bibinfo{person}{Ling Huang},
  \bibinfo{person}{Prateek Mittal}, \bibinfo{person}{Emil Stefanov},
  \bibinfo{person}{Vyas Sekar}, {and} \bibinfo{person}{Dawn Song}.}
  \bibinfo{year}{2012}\natexlab{}.
\newblock \showarticletitle{Evolution of social-attribute networks:
  Measurements, modeling, and implications using {Google+}}. In
  \bibinfo{booktitle}{\emph{Proceedings of the ACM Internet Measurement
  Conference}}. \bibinfo{pages}{131--144}.
\newblock


\bibitem[\protect\citeauthoryear{Gupta, Goel, Lin, and Sharma}{Gupta
  et~al\mbox{.}}{2013}]%
        {Gupta2013}
\bibfield{author}{\bibinfo{person}{Pankaj Gupta}, \bibinfo{person}{Ashish
  Goel}, \bibinfo{person}{Jimmy Lin}, {and} \bibinfo{person}{Aneesh Sharma}.}
  \bibinfo{year}{2013}\natexlab{}.
\newblock \showarticletitle{{WTF: The who to follow service at Twitter}}. In
  \bibinfo{booktitle}{\emph{Proceedings of the 22nd International Conference on
  World Wide Web}}. \bibinfo{pages}{505--514}.
\newblock


\bibitem[\protect\citeauthoryear{Hogg and Lerman}{Hogg and Lerman}{2012}]%
        {hogg2012social}
\bibfield{author}{\bibinfo{person}{Tad Hogg} {and} \bibinfo{person}{Kristina
  Lerman}.} \bibinfo{year}{2012}\natexlab{}.
\newblock \showarticletitle{Social dynamics of {Digg}}.
\newblock \bibinfo{journal}{\emph{EPJ Data Science}} \bibinfo{volume}{1},
  \bibinfo{number}{1} (\bibinfo{year}{2012}), \bibinfo{pages}{5}.
\newblock


\bibitem[\protect\citeauthoryear{Kossinets}{Kossinets}{2006}]%
        {Kossinets2006a}
\bibfield{author}{\bibinfo{person}{Gueorgi Kossinets}.}
  \bibinfo{year}{2006}\natexlab{}.
\newblock \showarticletitle{{Effects of missing data in social networks}}.
\newblock \bibinfo{journal}{\emph{Social Networks}} \bibinfo{volume}{28},
  \bibinfo{number}{3} (\bibinfo{year}{2006}), \bibinfo{pages}{247--268}.
\newblock


\bibitem[\protect\citeauthoryear{Leskovec, Backstrom, Kumar, and
  Tomkins}{Leskovec et~al\mbox{.}}{2008}]%
        {Leskovec2008}
\bibfield{author}{\bibinfo{person}{Jure Leskovec}, \bibinfo{person}{Lars
  Backstrom}, \bibinfo{person}{Ravi Kumar}, {and} \bibinfo{person}{Andrew
  Tomkins}.} \bibinfo{year}{2008}\natexlab{}.
\newblock \showarticletitle{{Microscopic evolution of social networks}}. In
  \bibinfo{booktitle}{\emph{Proceedings of the 14th ACM SIGKDD International
  Conference on Knowledge Discovery and Data Mining}}.
  \bibinfo{pages}{462--470}.
\newblock


\bibitem[\protect\citeauthoryear{Liben-Nowell and Kleinberg}{Liben-Nowell and
  Kleinberg}{2007}]%
        {liben2007link}
\bibfield{author}{\bibinfo{person}{David Liben-Nowell} {and}
  \bibinfo{person}{Jon Kleinberg}.} \bibinfo{year}{2007}\natexlab{}.
\newblock \showarticletitle{The link-prediction problem for social networks}.
\newblock \bibinfo{journal}{\emph{Journal of the Association for Information
  Science and Technology}} \bibinfo{volume}{58}, \bibinfo{number}{7}
  (\bibinfo{year}{2007}), \bibinfo{pages}{1019--1031}.
\newblock


\bibitem[\protect\citeauthoryear{Liu, He, Kapoor, and Srivastava}{Liu
  et~al\mbox{.}}{2013}]%
        {Liu2013c}
\bibfield{author}{\bibinfo{person}{Zhen Liu}, \bibinfo{person}{Jia~Lin He},
  \bibinfo{person}{Komal Kapoor}, {and} \bibinfo{person}{Jaideep Srivastava}.}
  \bibinfo{year}{2013}\natexlab{}.
\newblock \showarticletitle{{Correlations between community structure and link
  formation in complex networks}}.
\newblock \bibinfo{journal}{\emph{PLoS ONE}} \bibinfo{volume}{8},
  \bibinfo{number}{9} (\bibinfo{year}{2013}), \bibinfo{pages}{e72908}.
\newblock


\bibitem[\protect\citeauthoryear{L{\"{u}} and Zhou}{L{\"{u}} and Zhou}{2011}]%
        {Lu2011}
\bibfield{author}{\bibinfo{person}{Linyuan L{\"{u}}} {and} \bibinfo{person}{Tao
  Zhou}.} \bibinfo{year}{2011}\natexlab{}.
\newblock \showarticletitle{{Link prediction in complex networks: A survey}}.
\newblock \bibinfo{journal}{\emph{Physica A}} \bibinfo{volume}{390},
  \bibinfo{number}{6} (\bibinfo{year}{2011}), \bibinfo{pages}{1150--1170}.
\newblock


\bibitem[\protect\citeauthoryear{Mislove, Koppula, Gummadi, Druschel, and
  Bhattacharjee}{Mislove et~al\mbox{.}}{2008}]%
        {mislove-2008-flickr}
\bibfield{author}{\bibinfo{person}{Alan Mislove}, \bibinfo{person}{Hema~Swetha
  Koppula}, \bibinfo{person}{Krishna~P. Gummadi}, \bibinfo{person}{Peter
  Druschel}, {and} \bibinfo{person}{Bobby Bhattacharjee}.}
  \bibinfo{year}{2008}\natexlab{}.
\newblock \showarticletitle{Growth of the Flickr social network}. In
  \bibinfo{booktitle}{\emph{Proceedings of the 1st ACM SIGCOMM Workshop on
  Social Networks}}.
\newblock


\bibitem[\protect\citeauthoryear{Mitzenmacher}{Mitzenmacher}{2004}]%
        {mitzenmacher2004brief}
\bibfield{author}{\bibinfo{person}{Michael Mitzenmacher}.}
  \bibinfo{year}{2004}\natexlab{}.
\newblock \showarticletitle{A brief history of generative models for power law
  and lognormal distributions}.
\newblock \bibinfo{journal}{\emph{Internet Mathematics}} \bibinfo{volume}{1},
  \bibinfo{number}{2} (\bibinfo{year}{2004}), \bibinfo{pages}{226--251}.
\newblock


\bibitem[\protect\citeauthoryear{Newman}{Newman}{2001}]%
        {newman2001clustering}
\bibfield{author}{\bibinfo{person}{Mark Newman}.}
  \bibinfo{year}{2001}\natexlab{}.
\newblock \showarticletitle{Clustering and preferential attachment in growing
  networks}.
\newblock \bibinfo{journal}{\emph{Physical Review E}} \bibinfo{volume}{64},
  \bibinfo{number}{2} (\bibinfo{year}{2001}), \bibinfo{pages}{025102}.
\newblock


\bibitem[\protect\citeauthoryear{Schall}{Schall}{2015}]%
        {schall2015social}
\bibfield{author}{\bibinfo{person}{Daniel Schall}.}
  \bibinfo{year}{2015}\natexlab{}.
\newblock \bibinfo{booktitle}{\emph{Social network-based recommender systems}}.
\newblock \bibinfo{publisher}{Springer}.
\newblock


\bibitem[\protect\citeauthoryear{Viswanath, Mislove, Cha, and
  Gummadi}{Viswanath et~al\mbox{.}}{2009}]%
        {viswanath2009evolution}
\bibfield{author}{\bibinfo{person}{Bimal Viswanath}, \bibinfo{person}{Alan
  Mislove}, \bibinfo{person}{Meeyoung Cha}, {and} \bibinfo{person}{Krishna~P
  Gummadi}.} \bibinfo{year}{2009}\natexlab{}.
\newblock \showarticletitle{On the evolution of user interaction in facebook}.
  In \bibinfo{booktitle}{\emph{Proceedings of the 2nd ACM Workshop on Online
  Social Networks}}. \bibinfo{pages}{37--42}.
\newblock


\bibitem[\protect\citeauthoryear{Watts and Strogatz}{Watts and
  Strogatz}{1998}]%
        {watts1998collective}
\bibfield{author}{\bibinfo{person}{Duncan~J. Watts} {and}
  \bibinfo{person}{Steven~H. Strogatz}.} \bibinfo{year}{1998}\natexlab{}.
\newblock \showarticletitle{Collective dynamics of `small-world' networks}.
\newblock \bibinfo{journal}{\emph{Nature}} \bibinfo{volume}{393},
  \bibinfo{number}{6684} (\bibinfo{year}{1998}), \bibinfo{pages}{440}.
\newblock


\bibitem[\protect\citeauthoryear{Wu, Lin, Wang, and Gregory}{Wu
  et~al\mbox{.}}{2016}]%
        {wu2016link}
\bibfield{author}{\bibinfo{person}{Zhihao Wu}, \bibinfo{person}{Youfang Lin},
  \bibinfo{person}{Jing Wang}, {and} \bibinfo{person}{Steve Gregory}.}
  \bibinfo{year}{2016}\natexlab{}.
\newblock \showarticletitle{Link prediction with node clustering coefficient}.
\newblock \bibinfo{journal}{\emph{Physica A: Statistical Mechanics and its
  Applications}}  \bibinfo{volume}{452} (\bibinfo{year}{2016}),
  \bibinfo{pages}{1--8}.
\newblock


\bibitem[\protect\citeauthoryear{Yang, Lichtenwalter, and Chawla}{Yang
  et~al\mbox{.}}{2015}]%
        {Yang2015}
\bibfield{author}{\bibinfo{person}{Yang Yang}, \bibinfo{person}{Ryan~N.
  Lichtenwalter}, {and} \bibinfo{person}{Nitesh~V. Chawla}.}
  \bibinfo{year}{2015}\natexlab{}.
\newblock \showarticletitle{{Evaluating link prediction methods}}.
\newblock \bibinfo{journal}{\emph{Knowledge and Information Systems}}
  \bibinfo{volume}{45}, \bibinfo{number}{3} (\bibinfo{year}{2015}),
  \bibinfo{pages}{751--782}.
\newblock
\showISBNx{0219-1377, 0219-3116}
\showISSN{0219-1377}
\showeprint[arxiv]{1505.04094}


\bibitem[\protect\citeauthoryear{Yin, Gupta, Weninger, and Han}{Yin
  et~al\mbox{.}}{2010}]%
        {Yin2010}
\bibfield{author}{\bibinfo{person}{Zhijun Yin}, \bibinfo{person}{Manish Gupta},
  \bibinfo{person}{Tim Weninger}, {and} \bibinfo{person}{Jiawei Han}.}
  \bibinfo{year}{2010}\natexlab{}.
\newblock \showarticletitle{{A unified framework for link recommendation using
  random walks}}. In \bibinfo{booktitle}{\emph{Proceedings of the International
  Conference on Advances in Social Networks Analysis and Mining}}.
  \bibinfo{pages}{152--159}.
\newblock


\end{thebibliography}
\end{document}